\documentclass[usenatbib]{mn2e}

\usepackage{epsfig,color}
\usepackage{graphicx}
\usepackage{aas_macros}
\usepackage{lscape}

\title[Optical and $\gamma$-ray variability in blazars]{Connection between optical and $\gamma$-ray variability in blazars}
\author[T. Hovatta et al.]
{\parbox{\textwidth}{T. Hovatta$^{1}$\thanks{E-mail: thovatta@caltech.edu},
V. Pavlidou$^{2,3}$, 
O. G. King$^{1}$,
A. Mahabal$^{4}$,
B. Sesar$^{4}$,
R. Dancikova$^{4}$,
S. G. Djorgovski$^{4}$,
A. Drake$^{4}$,
R. Laher$^{5}$,
D. Levitan$^{4}$,
W. Max-Moerbeck$^{6}$,
E. O. Ofek$^{7}$,
T. J. Pearson$^{1}$,
T. A. Prince$^{4}$,
A. C. S. Readhead$^{1}$,
J. L. Richards$^{8}$ and
J. Surace$^{5}$}\vspace{0.4cm}\\
\parbox{\textwidth}{$^{1}$Cahill Center for Astronomy and Astrophysics, California Institute of Technology, Pasadena CA, 91125, USA\\
$^{2}$Physics Department, University of Crete, PO Box 2208, 71003 Heraklion, Greece\\
$^{3}$IESL, Foundation for Research and Technology-Hellas, PO Box 1527, 71110 Heraklion, Crete, Greece \\
$^{4}$Division of Physics, Mathematics, and Astronomy, California Institute of Technology, Pasadena, CA 91125, USA.\\
$^{5}$Spitzer Science Center, MS 314-6, California Institute of Technology, Pasadena, CA 91125, USA.\\
$^{6}$National Radio Astronomy Observatory, P.O. Box 0, Socorro, NM 87801, USA.\\
$^{7}$Benoziyo Center for Astrophysics, Faculty of Physics, Weizmann Institute of Science, Rehovot 76100, Israel.\\
$^{8}$Department of Physics, Purdue University, 525 Northwestern Ave, West Lafayette, IN 47907, USA.\\
}}
\begin{document}

\date{Accepted XXX. Received YYY; in original form ZZZ}

\pagerange{\pageref{firstpage}--\pageref{lastpage}} \pubyear{2013}

\maketitle

\label{firstpage}

\begin{abstract}
We use optical data from the Palomar Transient Factory (PTF) and the Catalina Real-Time Transient Survey (CRTS) to study the variability of $\gamma$-ray detected and non-detected objects in a large population of active galactic nuclei (AGN) selected from the Candidate Gamma-Ray Blazar Survey and Fermi Gamma-Ray Space Telescope catalogs. Our samples include 714 sources with PTF data and 1244 sources with CRTS data. We calculate the intrinsic modulation index to quantify the optical variability amplitude in these samples. We find the $\gamma$-ray detected objects to be more variable than the non-detected ones. The flat spectrum radio quasars (FSRQs) are more variable than the BL~Lac objects in our sample,  but the significance of the difference depends on the sample used. When dividing the objects based on their synchrotron peak frequency, we find the low synchrotron peaked (LSP) objects to be significantly more variable than the high synchrotron peaked (HSP) ones, explaining the difference between the FSRQs and BL~Lacs. This could be due to the LSPs being observed near their electron energy peak, while in the HSPs the emission is caused by lower energy electrons, which cool more slowly. We also find a significant correlation between the optical and $\gamma$-ray fluxes that is stronger in the HSP BL~Lacs than in the FSRQs. The FSRQs in our sample are also more Compton dominated than the HSP BL~Lacs. These findings are consistent with models where the $\gamma$-ray emission of HSP objects is produced by the synchrotron self-Compton mechanism, while the LSP objects need an additional external Compton component that increases the scatter in the flux-flux correlation. 
\end{abstract}


\section{Introduction}
The extra-galactic $\gamma$-ray sky is dominated by active galactic nuclei (AGN). The spectral energy distribution (SED) of AGN can be described by two components, 
a low energy component from radio to X-rays and a high energy one from X-rays to very high energy $\gamma$-rays. The low energy component can be attributed to
synchrotron radiation in a relativistic jet while the high energy emission can be either inverse Compton (IC) scattering of low-energy seed photons by the
synchrotron emitting electrons or emission through a hadronic process. If the seed photons for the IC scattering are the synchrotron photons, the process is called synchrotron self-Compton (SSC) \citep[e.g.][]{maraschi92,bloom96}. Alternatively, the seed photons can be external to the jet, for example, from the broad line region or the molecular torus, in which case the process is called external Compton (EC) \citep[e.g.][]{dermer93,sikora94}. For a recent study of the SED modelling of AGN, see \cite{bottcher13}.

If an AGN is viewed with its jet very close to the line of sight, it is called a blazar. Blazars emit brightly over the entire electromagnetic spectrum, and
can be further divided based on their optical classification into flat spectrum radio quasars (FSRQs) and BL~Lac objects. The optical emission of BL~Lac objects is dominated by a strong continuum and they show only weak emission lines. FSRQs have strong emission lines and are thought to have a much denser environment near the black hole. The two classes also differ in their large-scale jet properties. FSRQs are thought to be the beamed counterparts of Fanaroff-Riley type II (FR~II) galaxies, which have more powerful jets than the Fanaroff-Riley type I (FR~I) galaxies, thought to be the unbeamed counterpart of BL~Lac objects \citep{fanaroff74}. However, some recent studies have shown that the division of FSRQs and BL~Lacs into FR I and FR~II galaxies does not always follow this trend \citep[e.g.,][]{landt08,kharb10}.

Early studies using  the Energetic Gamma-Ray Experiment Telescope (EGRET) $\gamma$-ray instrument showed that high states in the optical and $\gamma$ rays are connected \citep[e.g.,][]{wagner95c, wagner95b, bloom97,
hartman01}, but detailed comparisons were hindered by the poor sampling of the light curves. The tendency for simultaneous flaring has been confirmed since the
launch of the Fermi Gamma-Ray Space Telescope (hereafter, {\it Fermi}) in 2008 and the capability of its Large Area Telescope (LAT) \citep{atwood09} to detect AGN even in their non-flaring
states. Many individual sources studied in great detail have shown correlated flares in optical and $\gamma$-ray spectral regions \citep[e.g.,][]{bonning09, marscher10, abdo10, agudo11,
ackermann12}. 

While individual sources have been studied in detail, the number of studies including multiple sources is limited. \cite{chatterjee12} studied the 6 best sampled
blazars observed within the SMARTS blazar monitoring program and found short delays between the optical and $\gamma$-ray flares. By decomposing the optical and
$\gamma$-ray light curves in to individual flares they also found the shapes of the flares to be similar in the two bands. However, \cite{bonning12} studied 12 sources in the same program and found that in some sources the variations were correlated while in others they were not. 

A statistical approach was taken by \cite{arshakian12} who studied 80 LAT-detected objects in the MOJAVE sample. Using non-simultaneous optical data, they found the
optical and $\gamma$-ray luminosities to be correlated in the FSRQs in their sample, while no significant correlation was found for
BL~Lac objects. The dispersion in the optical to $\gamma$-ray luminosity correlation was also much larger than for radio and $\gamma$-ray luminosities.
They attributed this to the possible larger variability in the optical compared to radio, which would increase the scatter in non-simultaneous correlations. 

We use the data for 714 AGN in the Palomar Transient Factory (PTF) \citep{rau09} and 1244 AGN in the Catalina Real-Time Transient Survey (CRTS) \citep{drake09} to study the connection between the optical and $\gamma$-ray emission in a {\it population} of AGN. The data sets have 637 sources in common. We use the intrinsic modulation index \citep{richards11} to study differences between sub-populations of sources and the significance estimation method of \cite{pavlidou12} to evaluate the flux -- flux correlation between the optical and $\gamma$-ray data. 

Our paper is organized as follows. In Sect.~\ref{sect:data} we describe our sample selection and data reduction. Sections \ref{sect:modindex} and \ref{sect:flux} show the results of our analysis. We discuss our results in the context of AGN models in Sect.~\ref{sect:discussion} and summarize our conclusions in Sect.~\ref{sect:conclusions}

\section{Sample and data reduction}\label{sect:data}
We use the AGN sample defined by \citep{richards11}  and monitored by them at 15\,GHz at the Owens Valley Radio Observatory (OVRO) 40-m as the starting point for our analysis. This sample of 1771 objects includes 1158 sources from the Candidate Gamma-ray Blazar Survey (CGRaBS) \citep{healey08} and all the sources above declination $-20^\circ$ that have been detected by the LAT in the first and second AGN catalogs \citep{abdo10b,ackermann11}. The OVRO sample also includes a small number of objects with interesting jet properties or that are being monitored by other programs. 

The CGRaBS sample is a statistically well-defined sample that was selected to resemble blazars that were detected by EGRET. The sources were selected from a flat-spectrum parent sample that is complete to 65\,mJy flux density at 4.8\,GHz and have radio spectral index\footnote{here and throughout the paper we define the spectral index as $S \propto \nu^\alpha$} $\alpha > -0.5$. The CGRaBS sources were then selected based on their radio spectral index, 8.4\,GHz flux density, and X-ray flux from {\it ROSAT} All Sky Survey. The sample was compiled before the launch of {\it Fermi} and was expected to contain a large number of sources that would be detected by {\it Fermi}. However, {\it Fermi} is much more sensitive to hard $\gamma$-ray spectrum sources than EGRET and a large number of CGRaBS sources have not been detected by {\it Fermi}. This makes the sample ideal for studying the differences between the {\it Fermi}-detected and non-detected objects. Our PTF and CRTS  samples contain 508 and 839 CGRaBS sources, respectively.  We emphasize that while our samples do not include all CGRaBS sources, they are {\it unbiased} subsets of the complete sample as the sample selection was not done based on the optical or $\gamma$-ray properties of the sources.

We search for all the objects within $1.5''$ of our sample targets in the PTF and $3''$ in the CRTS. These limits were selected based on the pixel size and typical seeing of the observations. We describe the details of the data extraction and analysis for each survey individually below. To allow for comparison with simultaneous OVRO and {\it Fermi} observations, we use data from 2008 to 2013. The redshifts for the sources in our sample range from 0.04 to 3.9 with a median of 1. For the population studies we divide our final samples into $\gamma$-ray loud and quiet objects based on their LAT detection. We use only the sources in the CLEAN samples of the 1st and 2nd LAT AGN catalogs \citep{abdo10,ackermann11}. These are sources for which only a single association has been determined in the {\it Fermi} catalogs. 

Additionally, we study the differences between the FSRQ and BL~Lac objects. The classifications we use come mainly from the CGRaBS and {\it Fermi} catalogs. Furthermore, we divide the sources based on the frequency of their synchrotron peak $\nu_p$ into low synchrotron peaked (LSP), intermediate synchrotron peaked (ISP) and high synchrotron peak (HSP) objects. We use the classifications from the 2nd LAT AGN catalog where the LSP sources have $\nu_p < 10^{14}$~Hz, ISPs have $10^{14} < \nu_p < 10^{15}$ and HSPs have $\nu_p > 10^{15}$~Hz \citep{ackermann11}. The number of sources in each subsample is listed in Table~\ref{table:sample}.

\begin{table*}
\centering
\caption{Number of sources in the PTF and CRTS samples for the various subsamples used in the analysis. There are 637 sources in common between the two samples.}\label{table:sample}
\begin{tabular}{lrrrrrrrr}
\hline
Sample & All & $\gamma$-ray loud & $\gamma$-ray quiet & FSRQ & BL~Lac & LSP & ISP & HSP \\
\hline
PTF & 714 & 313 & 401 & 448 & 168 & 127 & 39 & 68 \\
CRTS & 1244 & 543 & 701 & 717 & 293 & 193 & 67 & 131 \\
\hline
\end{tabular}
\end{table*} 

\subsection{PTF data}
The Palomar Transient Factory is designed to observe
optical transients and variable sources. It uses the 48-inch Samuel Oschin
Telescope at Palomar Observatory with R- and g$^{\prime}$-band filters and
a wide-field camera having a 7.2 square degree field of view. For a detailed
description of the project and its primary science goals see
\cite{law09} and \cite{rau09}. We find PTF R-band data for 870 objects in our sample. The limiting magnitude is about 20.5. Due to the
nature of the PTF observations, some areas of the sky get
better coverage than others and therefore the number of data points
for each source depends on the sky position and varies from
less than 10 to a few hundred.

We use the PTF Photometric Pipeline to extract the magnitudes. Images are processed using ``standard'' reduction procedures, including de-biasing, flat-fielding, and astrometric calibration. Catalogs are generated using SExtractor \citep{bertin96}. The PTF data are photometrically calibrated against the SDSS catalog.
Photometric nights are used in order to calibrate the photometry all over the
PTF footprints to accuracy of 2\% \citep{ofek12a,ofek12b}.
In addition, we apply relative calibration to the photometry,
with typical precision of a few  millimagnitudes at the bright end (magnitude 15).
The relative photometry algorithm is described in \citet{ofek11}.

We discard data based on various flags given by the SExtractor (e.g., blending of multiple sources within the field, bad astrometry of the field) and additional flags given by the PTF Photometric Pipeline (e.g., saturated or dead pixels). The sources can be observed up to four times each night and we average the magnitudes over each night. This is important because our variability analysis method assumes subsequent observations to be independent and multiple observations within a night may introduce biases. We acknowledge that AGN can be variable on time scales less than a day \citep[e.g.,][]{wagner95} but the variability amplitude is typically a fraction of a magnitude.
For our further analysis we select all the objects that have at least three data points. By visually examining all the light curves, we also discarded sources which clearly suffered from blending even if the data were not flagged by SExtractor. This results in a sample of 714 sources which are listed in Table~\ref{table:PTF}.

We correct the PTF R-band magnitudes for Galactic extinction using the re-calibrated dust maps of \cite{schlafly11} with the reddening law of \cite{fitzpatrick99} extracted from NASA/IPAC Extragalactic Database (NED). The corrected magnitudes were then converted into flux density units (mJy) by using a zero point of 3631\,Jy. We note that there is an additional colour term correction that affects the conversion \citep{ofek12a} but it is less than 0.04\,mag for a typical blazar spectrum and therefore ignored. In all our further analysis we use the light curves in flux density units.

\subsection{CRTS data}
The Catalina Real-Time Transient Survey\footnote{http://crts.caltech.edu} uses the data from the Catalina Sky Survey (CSS)\footnote{http://www.lpl.arizona.edu/css/} to report on optical transients. Observations are made without a specific filter with the 0.68-m Catalina Schmidt Telescope in Arizona, USA, the 0.5-m Uppsala Schmidt at Siding Spring Observatory, NSW, Australia, and the 1.5-m reflector located on Mt. Lemmon in Arizona. The main science goal of CSS is to detect near-Earth objects but the large sky coverage makes it ideal for transient studies. For details of the CRTS and the first results see \cite{drake09}, \cite{mahabal11}, and \cite{djorgovski12}. We find CRTS data from the Catalina Surveys Data Release 2 \citep{drake09} for 1335 sources in our sample. The magnitudes for all the sources are derived using the SExtractor program. The limiting magnitude of CRTS is about 21.

We discard data using the blend flag of SExtractor which indicates if the source is blended with another one in the field. Because the pipeline is not optimized for AGN studies, there are many outliers in the data even after the basic flagging. Outliers have a significant effect on our modulation index analysis, so we discard the most extreme cases. All the sources are observed four times each night within $\sim$~30 min time period. We use this additional information to discard any data points which differ by more than 0.8 magnitudes from the other data points over the same night, and all the data over a night if the standard deviation of the data points exceeds 0.5~mag.  These limits were empirically determined from the data to exclude high outliers that are unlikely to be due to intra-night variability. In the same way as for the PTF data, we average all the observations within a single night. We then select all the sources with at least three data points which results in a sample of 1244 objects. All the sources are listed in Table~\ref{table:CRTS}.

For the modulation index study we must convert the magnitudes into flux density units. In the case of the CRTS data this is not straightforward because the observations are done without a filter. However, the setup used resembles a $V$-band magnitude and it has been empirically determined that the CRTS magnitudes can be converted into Cousins $V$-band by the relation\footnote{http://nesssi.cacr.caltech.edu/DataRelease/FAQ2.html} $V = V_{\rm CSS} + 0.91 \times (V-R)^2 + 0.04$ where $(V-R)$ depends on the spectrum of the object. Because the $(V-R)$ colour varies depending on the flux state of the object \cite[e.g.,][]{bonning12}, and is unknown for a large fraction of the sources in our sample, we take a statistical approach in the conversion. \cite{jester05}  derive an empirical relation $(V-R) = 0.38\times (r-i) + 0.27$ applicable for quasars at redshifts $<2$ using the SDSS data. We extract $(r-i)$ colours for 10~000 quasars using the SDSS DR5 sky server SQL tool\footnote{http://cas.sdss.org/dr5/en/proj/advanced/quasars/query.asp} to obtain a mean value of 0.34 for the $(V-R)$ of quasars. We use this value to convert the CSS magnitudes into Cousins $V$-band magnitudes, and a zero point of 3953 Jy to convert into mJy. There are 123 sources in our CRTS sample with redshift larger than 2, which are outside the nominal redshift range for the conversion. We note that finding the exact conversion factor is not critical because the multiplicative factor will cancel out in the calculation of the modulation index. However, this will have an effect on the flux-flux correlation and therefore we only use the PTF data in the correlation analysis.

\section{Intrinsic Modulation Index}\label{sect:modindex}
Variability amplitudes in light curves can be studied in numerous ways, for example using the variability index \citep[e.g.,][]{aller92}, the fractional variability amplitude \citep[e.g.,][]{edelson02}, and the modulation index \citep[e.g.,][]{kraus03}. While all these methods are widely used, they do not account for the effects of irregular sampling and are often applicable only when the variations significantly exceed the measurement errors. Furthermore, they do not provide an estimate of the error in the variability measure. 

The intrinsic modulation index $\overline{m}$ was first defined and used in \cite{richards11} on the OVRO 40-m 15 GHz data. 
Its basic principle is the same as the standard modulation index, defined as the standard deviation of the flux density measurements in units of the mean flux density. The main difference is that it is calculated using the intrinsic standard deviation  $\sigma_0$ and the intrinsic mean flux density $S_0$ of the observation so that
\begin{equation}
\centering
\overline{m} = \frac{\sigma_0}{S_0}. 
\end{equation}

The term intrinsic denotes values that we would obtain if we had zero observational errors and infinite number of samples. The intrinsic values are calculated using a likelihood approach which assumes the observed flux densities to follow a normal distribution with Gaussian errors. The measurement errors are accounted for in the calculation of the joint likelihood for $S_0$ and $\overline{m}$. For full derivation of the likelihoods see \cite{richards11}.

The main advantage of this method is that it is possible to determine an uncertainty for the modulation index which is needed in the comparison of the different source populations. For sources with $\overline{m}$ within $3\sigma$ of zero, the $3\sigma$ upper limit will be determined. Another advantage is that the number of data points in each light curve is accounted for in the errors of $\overline{m}$ so that for sources with fewer data points the errors will be larger. The minimum number of data points needed for the analysis is three. The intrinsic modulation indices and the intrinsic flux densities are listed for each source in the PTF sample in Table~\ref{table:PTF} and in Table~\ref{table:CRTS} for the CRTS sample.

\begin{figure*}
\includegraphics[scale=0.45]{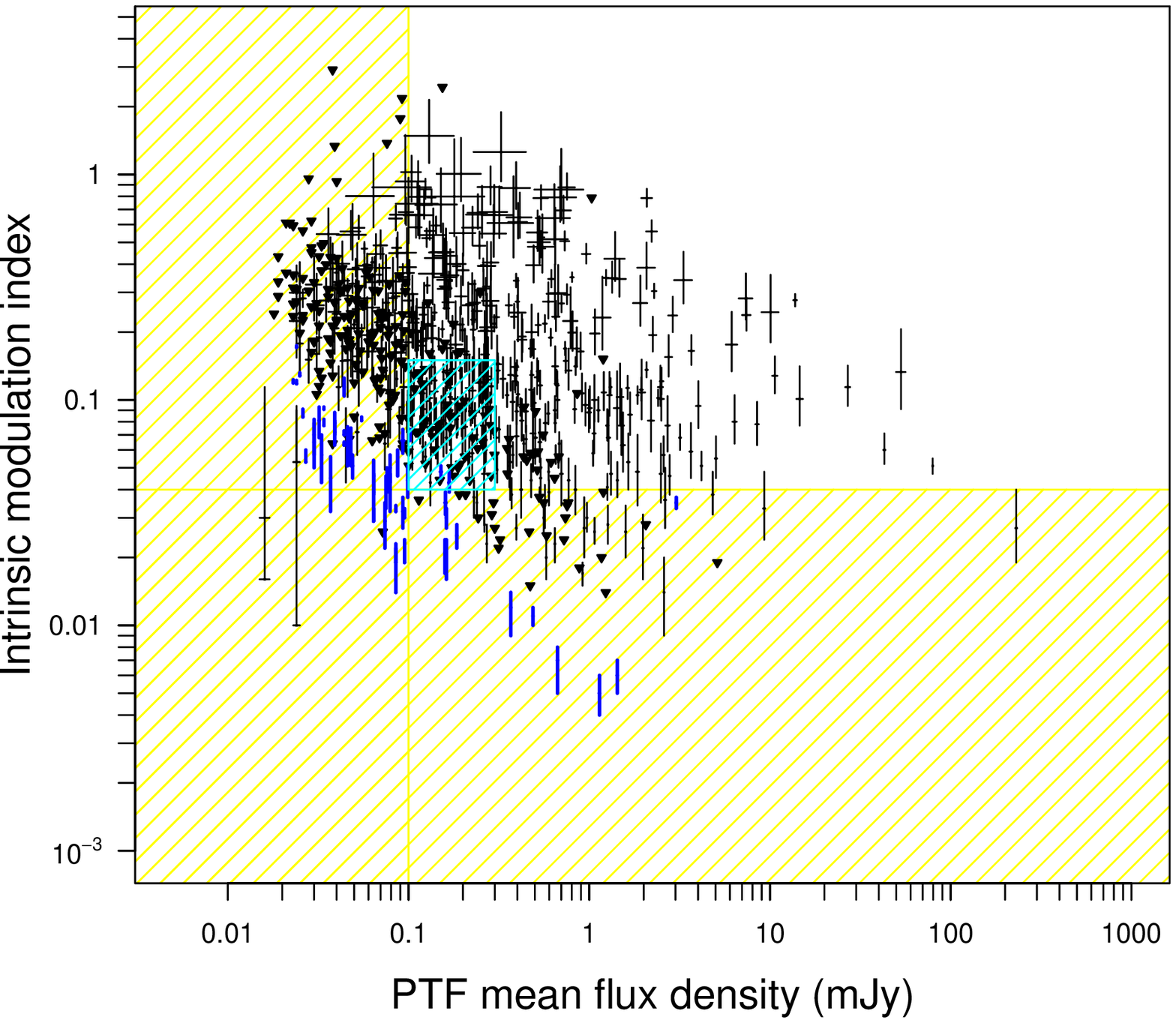}\includegraphics[scale=0.45]{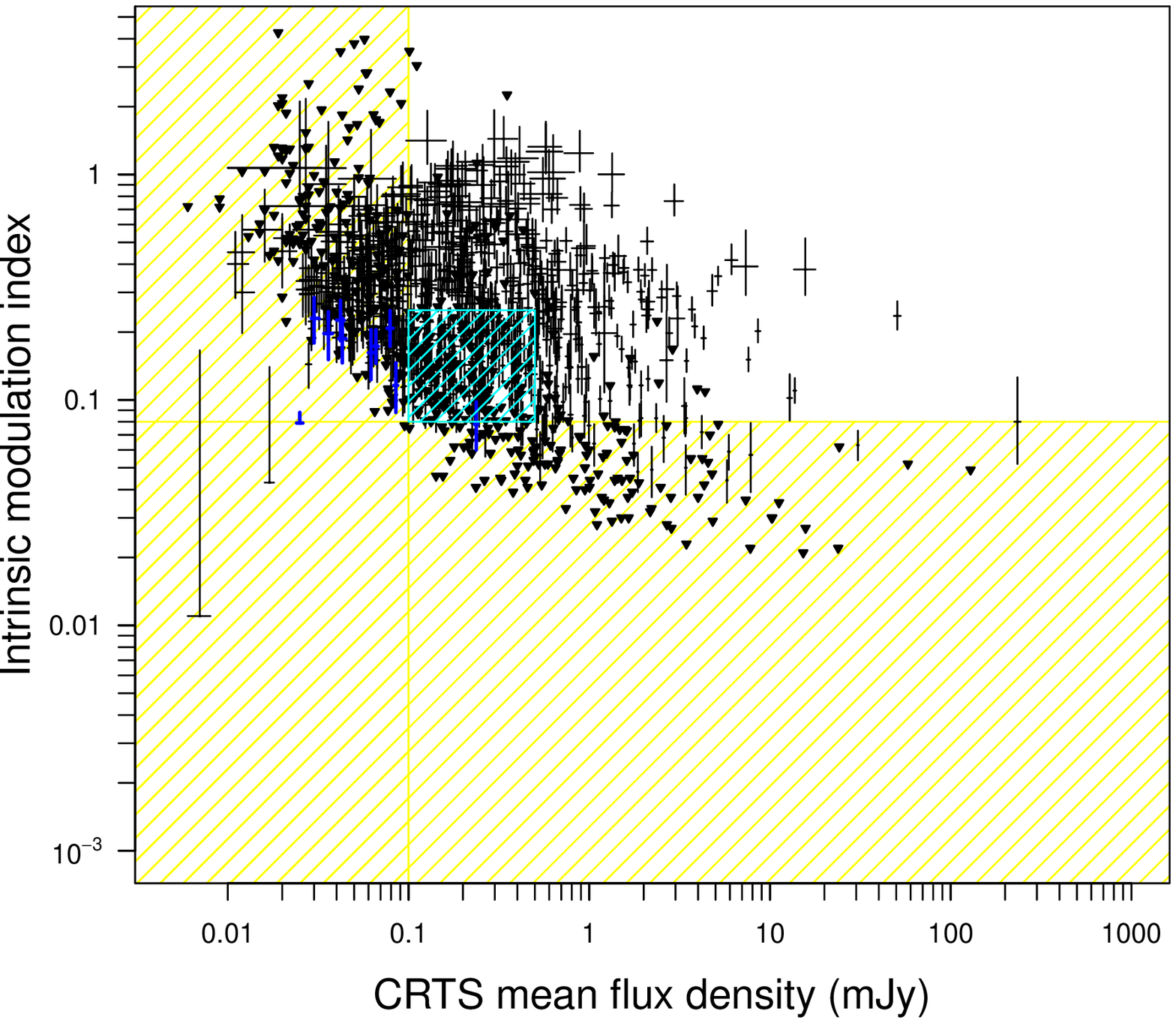}
\caption{Intrinsic modulation index $\overline{m}$ with 1~$\sigma$ uncertainty plotted against the intrinsic mean flux density $S_0$ of all the sources in the PTF (left) and CRTS (right) samples. Black triangles are 3~$\sigma$ upper limits of $\overline{m}$. Thick blue symbols are white dwarfs used as calibrators. Yellow and cyan hatched areas show the regions not included in the population studies. See the text for details.}\label{modindex}
\end{figure*}

There are some important limitations with the method and the data that are illustrated in Fig.~\ref{modindex}, which shows the intrinsic modulation index plotted against the intrinsic mean flux density of each object.  One limitation is the systematic errors in the data sets. It is often the case that sources that are supposed to be constant in flux density will show non-zero modulation index due to systematic uncertainties in the data. In order to test this effect on the PTF and CRTS data, we use data for white dwarfs, which are assumed to be constant in the optical band, to determine the limit above which the variations we detect are due to intrinsic changes in the object and not due to systematic uncertainties. In the PTF data, the highest intrinsic modulation index for a white dwarf at a flux density above 0.3\,mJy (see below for the justification of this cutoff) is $\overline{m}=0.035\pm0.002$ and therefore we set the lower limit of the intrinsic modulation index to 0.04 for PTF blazars. Similarly, in the CRTS data the highest value for a white dwarf at high fluxes is $\overline{m} = 0.078\pm0.02$ and we set the lower limit to 0.08 for the CRTS blazars. 

 For sources weaker than a certain flux density limit, the method gives mainly upper limits and shows that below the weak flux density limit, we cannot reliably determine the intrinsic modulation index. This limit is 0.1\,mJy for the PTF and CRTS samples, estimated such that we include as few marginal non-detections as possible (i.e. sources with very high upper limits on the intrinsic modulation index). This limit is also required because the white dwarfs show a dependency on the flux density so that at lower fluxes the limit for modulation index is higher, indicating that the errors in the data are underestimated at low fluxes, making the intrinsic modulation index estimates unreliable. These limits are shown in Fig.~\ref{modindex} as yellow hatched regions.

There is also a large concentration of upper limits in both plots at the lower left corner (cyan hatched region in Fig.~\ref{modindex}). This is a region where we cannot reliably determine the intrinsic modulation index for sources that are fainter than 0.3\,mJy in the PTF sample and 0.5\,mJy in the CRTS sample. This effect disappears when the modulation index is above 0.15 for the PTF sample and 0.25 for the CRTS sample. In the following population studies we will exclude the regions outside these limits (yellow and cyan hatched regions in Fig.~\ref{modindex}) to ensure that our results are not biased due to incomplete sampling in these regions.  We note that our analysis is not sensitive to the exact value of the limits, and we have tested that even 50\% higher flux density and modulation index limits would not change any of our conclusions. After these cutoffs and excluding the upper limits we have 271 sources in the PTF sample and 343 sources in the CRTS sample.

Additional caveats for the method are listed in \cite{richards11}. One of the most important is the assumption of Gaussian flux density distributions and the leakage of probability density to negative flux densities when this is not met. It is clear that the flux density distribution for many of our sources is non-Gaussian, especially when there are only a few data points. This results in long tails in the probability density distribution that in the case of high $\overline{m} \ge 0.5$ leak to unphysical negative flux densities. We estimate that for sources with $\overline{m} \ge 0.7$ about 8\% of the ``true flux density'' probability density leaks to negative values. For sources with $\overline{m} \ge 1.0$ this becomes $\sim17$\%. There are 21 sources in the PTF sample and 73 in the CRTS sample with $\overline{m} >= 0.7$. The solution to this problem is to extend the likelihood analysis to other shapes of flux density distributions and will be presented in a forthcoming publication.

\subsection{Redshift dependence}
\begin{figure*}
\includegraphics[scale=0.5]{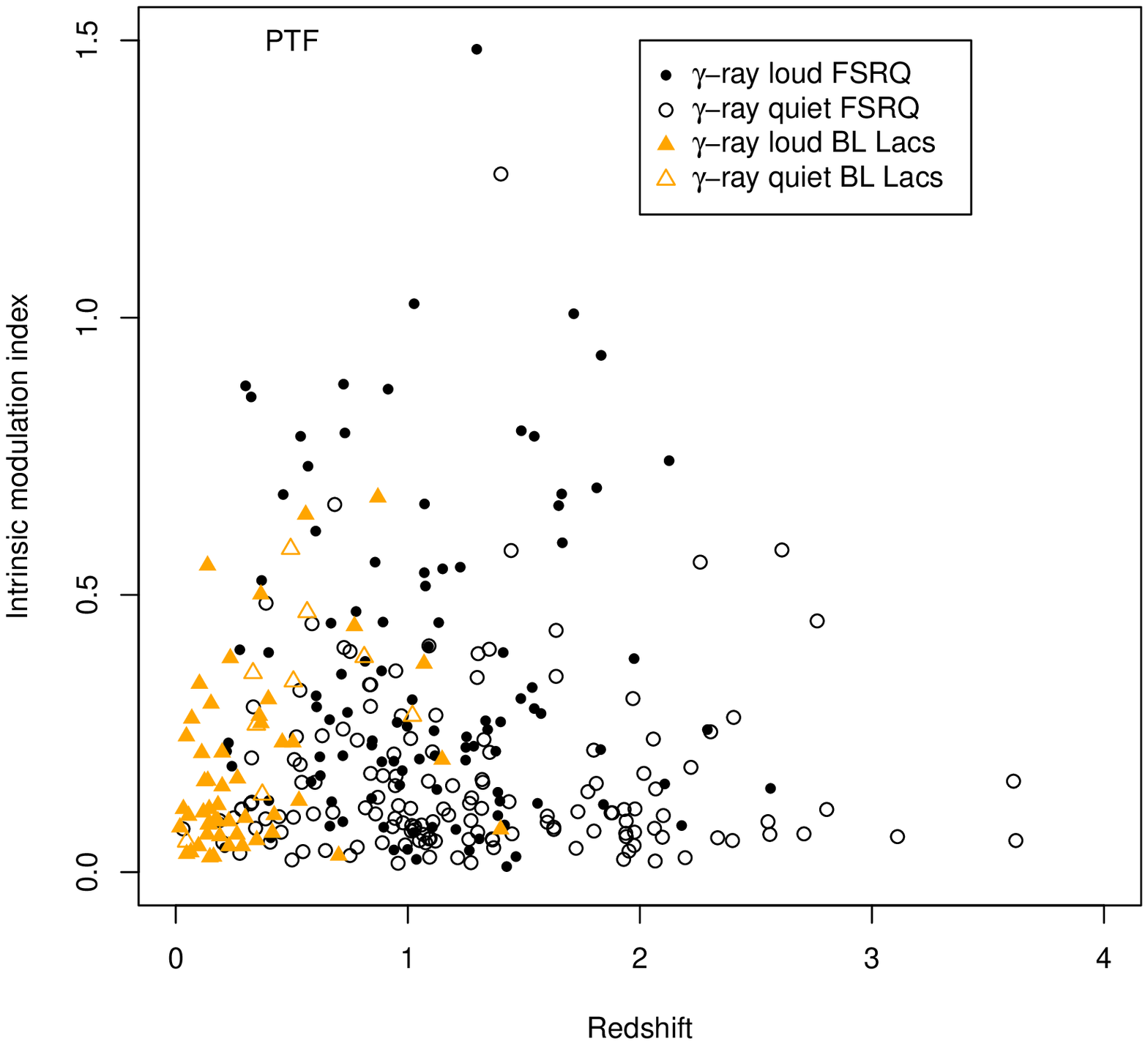}\includegraphics[scale=0.5]{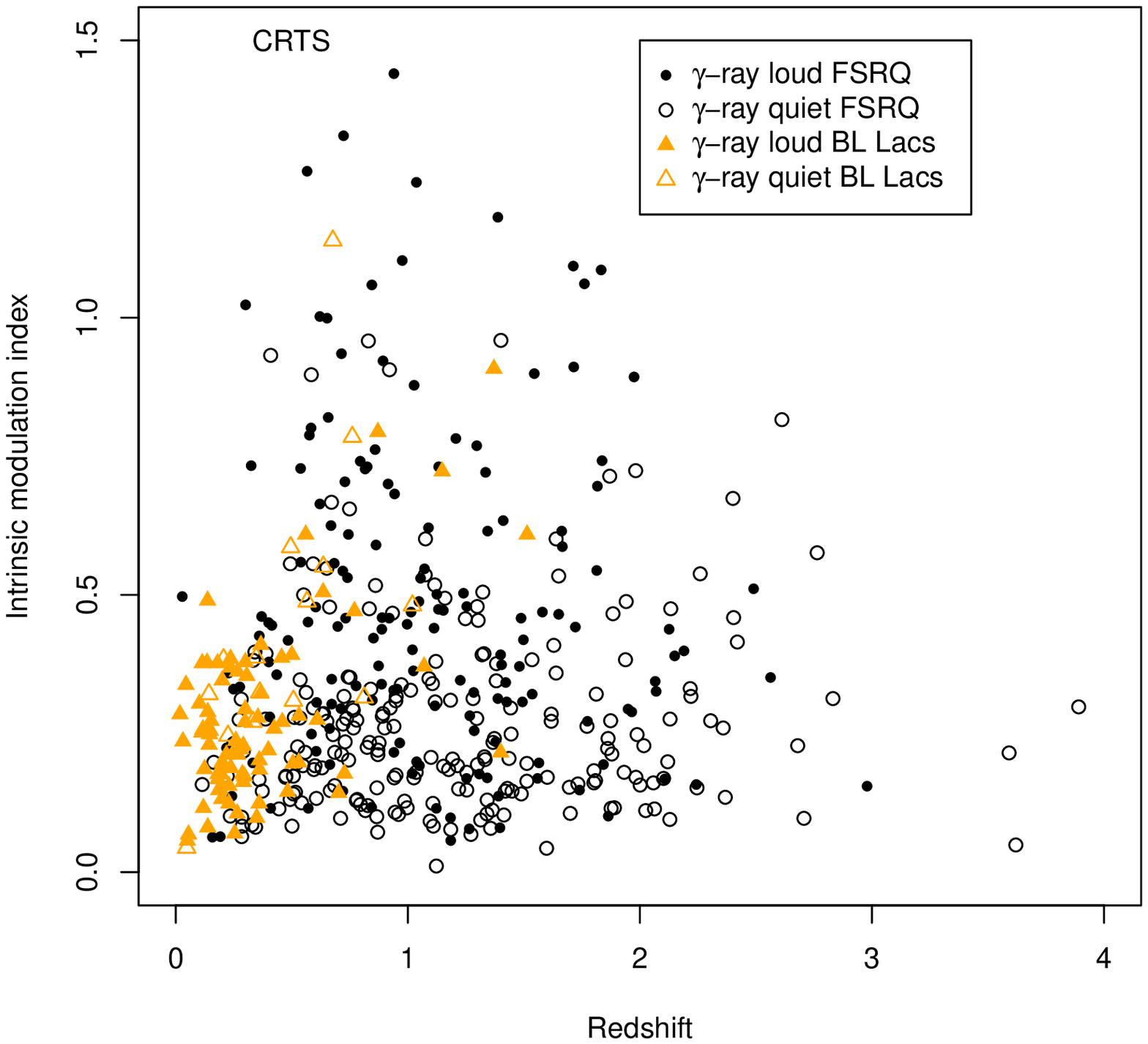}
\caption{Intrinsic modulation index against the redshift for the PTF (left) and CRTS (right) samples. {\it Fermi}-detected FSRQs are shown with black filled circles, non-detected FSRQs with open black circles, {\it Fermi}-detected BL~Lacs with filled orange triangles, and non-detected BL~Lacs with open orange triangles.}\label{redshift}
\end{figure*}

The redshift range for $\gamma$-ray-detected sources is limited due to the capability of {\it Fermi} to detect sources at high redshifts. Similarly, BL~Lac objects have typically lower redshifts than FSRQs. In Fig.~\ref{redshift}, we show $\overline{m}$ against the redshift for the {\it Fermi}-detected  and non-detected FSRQs and BL~Lacs in both PTF and CRTS samples. The redshift range for {\it Fermi}-detected sources extends to about $z=2$ and the redshift range of BL~Lacs to $z \sim 1$ while the non-detected sources and FSRQs have a range of redshifts up to almost 4.

We have searched for correlations between $\overline{m}$ and redshift using the non-parametric Kendall's tau test. FSRQs do not show any significant correlations in the data. In BL~Lacs there is no significant correlation if we limit the redshifts to $z<1$. Above this limit the redshifts of BL~Lacs are incomplete \citep[e.g.][]{shaw13} and the correlation is biased. As shown in Sect.~\ref{sect:optclass} below, the mean intrinsic modulation indices for FSRQs at $z<1$ and $z>1$ are similar to within $1\sigma$, showing that there is no redshift dependence. These results show that our conclusions are not affected by the different redshift range of the populations, but for completeness, in the following analysis of $\gamma$-ray detected and non-detected sources, and BL~Lacs and FSRQs, we show the results for both the total sample and samples restricted to a common redshift range.

\subsection{Mean intrinsic modulation index}
\begin{figure}
\includegraphics[scale=0.4]{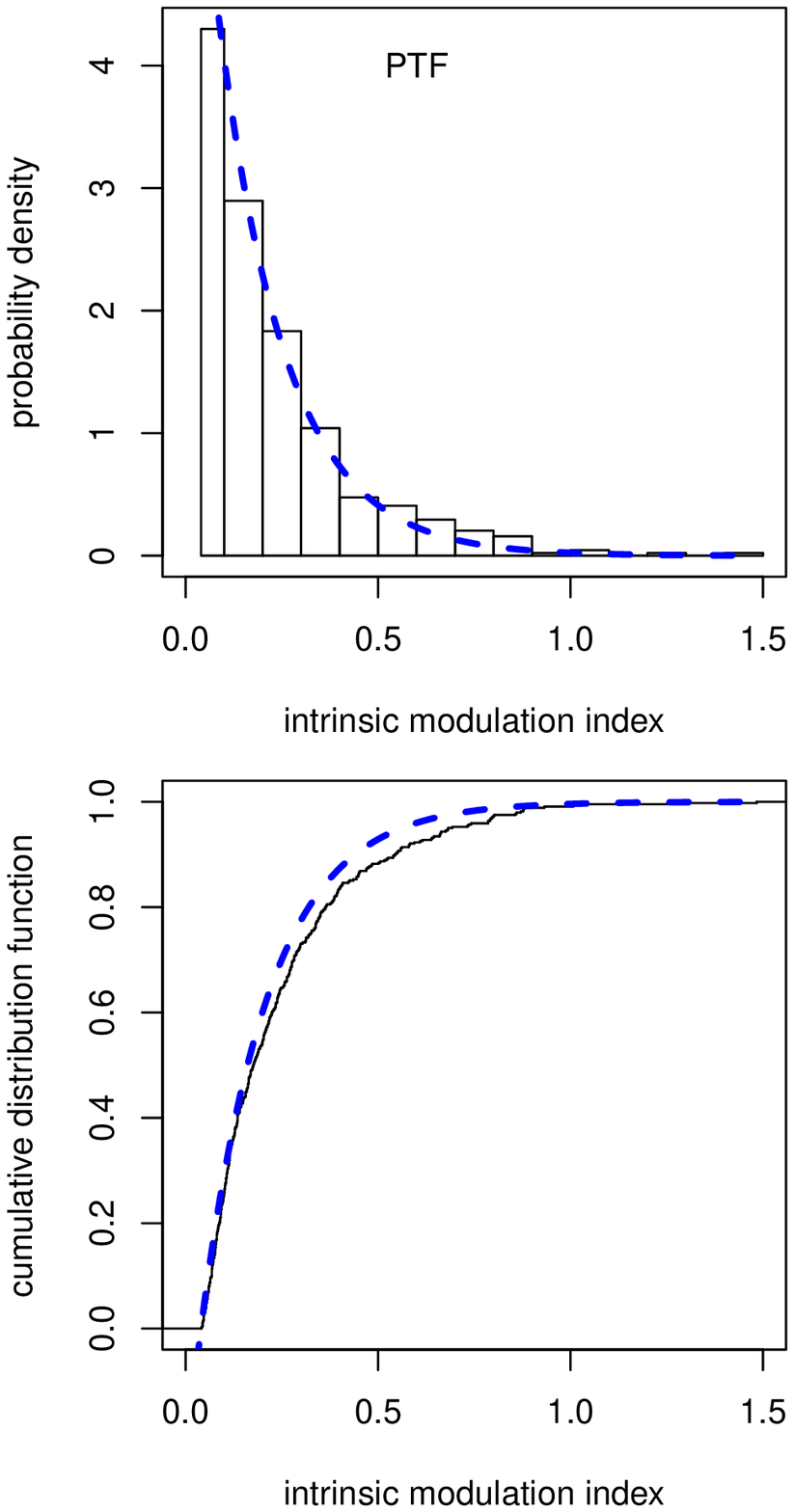}\includegraphics[scale=0.4]{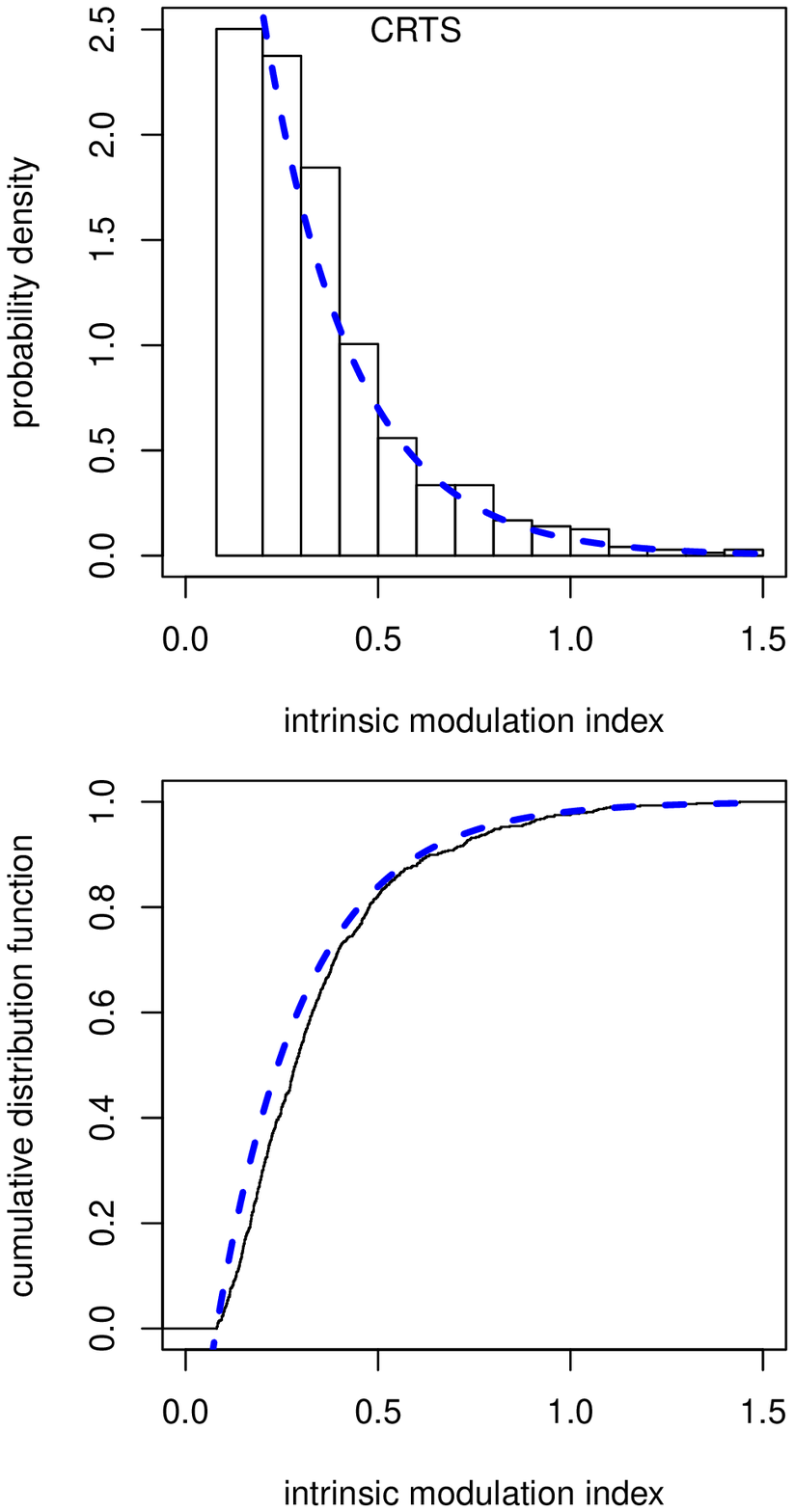}
\caption{Top: Histogram of the maximum-likelihood intrinsic modulation indices $\overline{m}$ for the PTF sources with $\overline{m} > 0.04$ (left) and for the CRTS sources with $\overline{m} > 0.08$, normalized as a probability density. The blue dashed line shows an exponential function with a mean 0.18 (PTF) and 0.23 (CRTS), where the mean values are determined using the maximum-likelihood analysis. Bottom: Same data as in the top panel but shown with a cumulative distribution function.}\label{fluxdist}
\end{figure}

In the following sections we will investigate whether the intrinsic modulation index correlates with various physical properties of the objects, such as the $\gamma$-ray detection, optical classification, and location of the synchrotron peak. We do this using a maximum likelihood approach which is described in \cite{richards11}.  As shown in Fig.~\ref{fluxdist} the distribution of $\overline{m}$ for the sources  approximately follows an exponential function. This allows us to calculate the probability density of the mean intrinsic modulation index $m_0$. The errors of $m_0$ are determined from the probability density distribution and they do not have to be symmetric. We can then compare the  means of the various sub-populations and determine whether they agree within some significance limit. We consider a result significant at the $3\sigma$ level if the p-value is less than 0.0027. The mean intrinsic modulation indices for each sub-population are listed in Table~\ref{table:modindex}. 

In order to study the difference between the populations, we calculated the likelihood of the difference between the mean intrinsic modulation indices in the two populations by taking the cross-correlation of the individual $m_0$ likelihoods for each population.  This allows us to estimate the most likely value for the difference between the populations. These are listed in Table~\ref{table:corr}. 

\begin{table*}
\centering
\caption{The mean intrinsic modulation indices and their 1$\sigma$ errors in the PTF and CRTS samples for the various sub-populations used in the analysis.}\label{table:modindex}
\begin{tabular}{lrrrrrrr}
\hline
Sample & $\gamma$-ray loud & $\gamma$-ray quiet & FSRQ & BL~Lac & LSP & ISP & HSP \\
\hline
PTF & $0.201^{+0.015}_{-0.014}$ & $0.096^{+0.015}_{-0.012}$ & $0.215^{+0.023}_{-0.020}$ & $0.149^{+0.015}_{-0.014}$ & $0.277^{+0.035}_{-0.030}$ & $0.199^{+0.038}_{-0.031}$ & $0.096^{+0.017}_{-0.014}$  \\
CRTS & $ 0.254^{+0.016}_{-0.015}$ & $0.163^{+0.022}_{-0.019}$ & $0.315^{+0.031}_{-0.027}$ & $0.179^{+0.015}_{-0.014}$ & $0.365^{+0.039}_{-0.035}$ & $0.243^{+0.041}_{-0.034}$ & $0.123^{+0.018}_{-0.015}$ \\
\hline
\end{tabular}
\end{table*} 

\begin{table*}
\centering
\caption{Most likely difference between the mean intrinsic modulation indices and their 1$\sigma$ errors and significances for the various sub-populations tested.}\label{table:corr}
\begin{tabular}{lrrrr}
\hline
Populations & PTF & Signif. & CRTS & Signif. \\
\hline
$\gamma$-ray loud vs. $\gamma$-ray quiet & $-0.103^{+0.020}_{-0.020}$ & $4.5\sigma$ & $-0.090^{+0.027}_{-0.025}$ & $3.0\sigma$ \\
CGRaBS $\gamma$-ray loud vs. quiet & $-0.116^{+0.030}_{-0.032}$ & $>5\sigma$ & $-0.117^{+0.085}_{-0.073}$ & $4.1\sigma$ \\
FSRQ vs. BL~Lac &  $-0.065^{+0.025}_{-0.027}$ & $2.6\sigma$&  $-0.135^{+0.031}_{-0.034}$ & $4.6\sigma$ \\
CGRaBS FSRQ vs. BL~Lac & $-0.011^{+0.041}_{-0.038}$ & $<1\sigma$ & $-0.068^{+0.043}_{-0.042}$ & $1.6\sigma$ \\
LSP vs. ISP & $-0.076^{+0.049}_{-0.047}$ & $1.5\sigma$ & $-0.120^{+0.054}_{-0.053}$ & $2.1\sigma$ \\
LSP vs. HSP & $-0.178^{+0.034}_{-0.037}$ & $>5\sigma$ & $-0.240^{+0.034}_{-0.037}$ & $>5\sigma$\\
ISP vs HSP & $-0.101^{+0.035}_{-0.041}$ & $3.1\sigma$ & $-0.119^{+0.038}_{-0.044}$ & $3.4\sigma$\\
\hline
\end{tabular}
\end{table*}

\subsection{$\gamma$-ray detected vs. non-detected sources}
\begin{figure}
\includegraphics[scale=0.6]{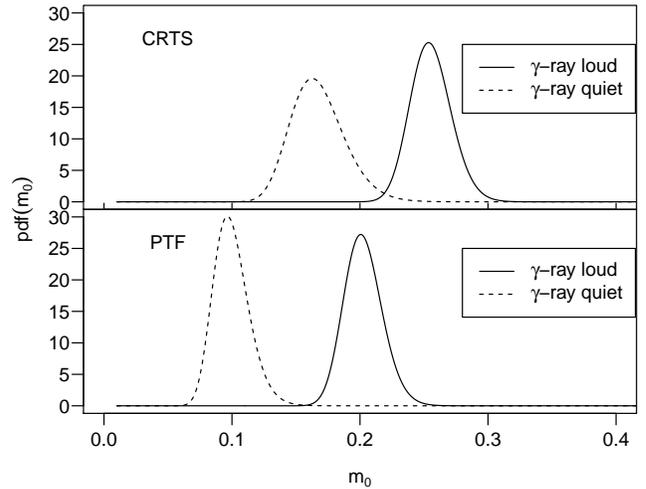}
\caption{Probability density of the  mean intrinsic modulation index for the $\gamma$-ray detected and non-detected objects in the CRTS (top) and PTF (bottom). Solid line is $\gamma$-ray loud objects. Dashed line is $\gamma$-ray quiet objects.}\label{gammadet}
\end{figure}

Figure~\ref{gammadet} shows the probability density of the  mean intrinsic modulation index for the $\gamma$-ray detected and non-detected objects in the CRTS (top) and PTF (bottom) samples. The two distributions are not consistent with a single value and the $\gamma$-ray detected objects are more variable than the non-detected ones. Both PTF and CRTS give consistent results. We find that the most likely difference between the mean intrinsic modulation indices is more than $3\sigma$ from zero. 

If we limit the likelihood analysis to objects with redshift $z < 2$ (the common range for the detected and non-detected samples), the difference between the means of the {\it Fermi}-detected ($m_{0,\rm{PTF}} = 0.230^{+0.023}_{-0.020}$ and $m_{0,\rm{CRTS}} = 0.302^{+0.026}_{-0.024}$) and non-detected ($m_{0,\rm{PTF}} = 0.072^{+0.013}_{-0.012}$ and $m_{0,\rm{CRTS}} = 0.161^{+0.027}_{-0.022}$) sources is still highly significant with the most likely difference of $-0.156^{+0.024}_{-0.048}$ for PTF and $-0.139^{+0.036}_{-0.035}$ for CRTS.

One caveat is that the sample we use is not statistically complete as it is a combination of the CGRaBS and {\it Fermi}-detected sources which have different selection criteria. Therefore we repeat the analysis using the CGRaBS sources only,  which is an unbiased sample. In this case the difference between the $\gamma$-ray detected and non-detected objects is even larger (see Table~\ref{table:corr}). This is mainly due to {\it Fermi} being more sensitive to HSP objects which are less variable than the LSP sources that dominate the CGRaBS sample, as we show below.

\subsection{FSRQs vs BL~Lac objects}\label{sect:optclass}
\begin{figure}
\includegraphics[scale=0.6]{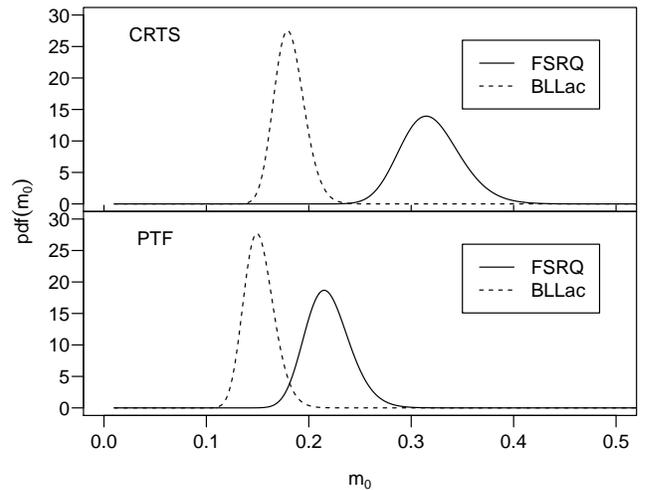}
\caption{Probability density of the mean intrinsic modulation index for FSRQs and BL~Lac objects in the CRTS (top) and PTF (bottom). Solid line is FSRQs, and dashed line is BL~Lac objects.}\label{optclass}
\end{figure}

Figure~\ref{optclass} shows the probability density of the  mean intrinsic modulation index for the FSRQs and BL~Lac objects in the CRTS (top) and PTF (bottom) samples. Again, the two distributions are not consistent with a single value and the FSRQs are more variable than BL~Lac objects. The most likely difference between the mean intrinsic modulation index is more than $3\sigma$ from zero for the CRTS sample and nearly $3\sigma$ from zero for the PTF sources.  If we repeat the analysis for the CGRaBS sources only, the most likely difference between the two populations is less than $1\sigma$ (see Table~\ref{table:corr}).

If we constrain our likelihood analysis to objects with $z < 1$, we find that the means of FSRQs ($m_{0,\rm{PTF}} = 0.197^{+0.030}_{-0.025}$ and $m_{0,\rm{CRTS}} = 0.318^{+0.041}_{-0.035}$) differ from the BL~Lacs ($m_{0,\rm{PTF}} = 0.152^{+0.026}_{-0.021}$ and $m_{0,\rm{CRTS}} = 0.189^{+0.027}_{-0.023}$) by less than $2\sigma$ in the PTF with the most likely difference of $-0.044^{+0.036}_{-0.037}$ and by about $2\sigma$ in the CRTS with the most likely difference of $-0.127^{+0.044}_{-0.047}$. The lower significance compared to the full sample is probably due to the smaller number of FSRQs used in the limited analysis. We test this by estimating the means of the FSRQs at $z < 1$ and at $z > 1$. The mean values at $z < 1$ are listed above and the most likely difference to the FSRQs at $z > 1$ ($m_{0,\rm{PTF}} = 0.235^{+0.037}_{-0.031}$ and $m_{0,\rm{CRTS}} = 0.309^{+0.050}_{-0.041}$) is less than $1\sigma$ for both the PTF sample with the most likely difference of $-0.038^{+0.043}_{-0.045}$ and the CRTS sample with the most likely difference of $-0.009^{+0.061}_{-0.059}$.

\subsection{Division based on SED classification}
\begin{figure}
\includegraphics[scale=0.6]{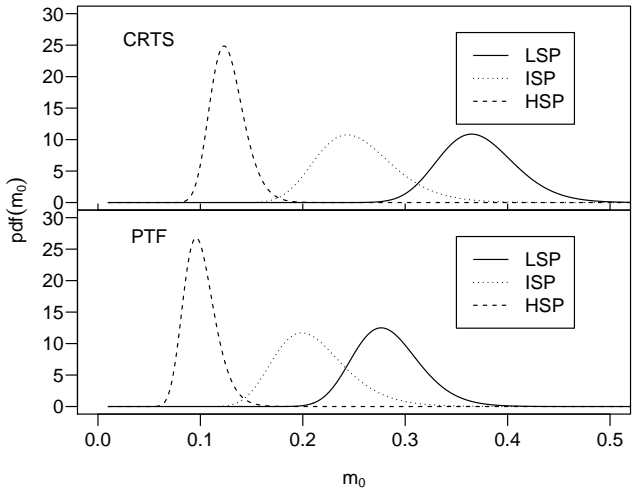}
\caption{Probability density of the mean intrinsic modulation index for LSP, ISP and HSP objects in the CRTS (top) and PTF (bottom). Solid line is LSPs, dotted line is ISP objects, and dashed line is HSP objects.}\label{sedclass}
\end{figure}

Figure~\ref{sedclass} shows the probability density of the  mean intrinsic modulation index for the LSP, ISP and HSP sources in the CRTS (top) and PTF (bottom) samples. There is a clear trend in both data sets for the LSP sources to be more variable than the ISP and HSP sources. In the case of LSP vs. ISP sources we cannot distinguish the mean intrinsic modulation indices from each other within $2\sigma$ but the difference between LSP and HSP sources is highly significant. The mean intrinsic modulation index difference is more than $3\sigma$ from zero for both PTF and CRTS samples.

\begin{figure}
\includegraphics[scale=0.4]{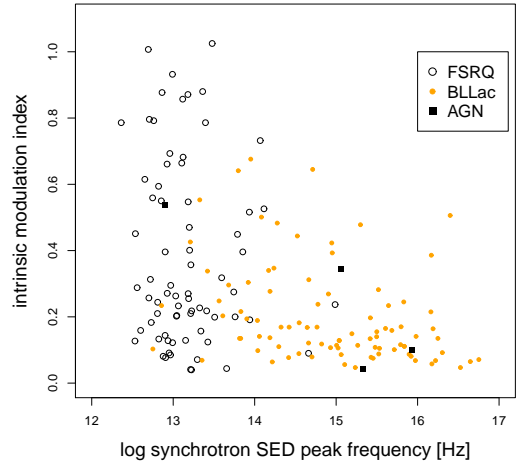}
\caption{Intrinsic modulation index from PTF plotted against the peak frequency of the synchrotron component. FSRQs are shown with open circles, BL~Lac objects with filled orange circles and AGN without optical classification with black squares.}\label{mod_nu}
\end{figure}

We also compare the intrinsic modulation index directly to the peak frequency of the synchrotron component. We use the peak frequencies used to determine the spectral classifications tabulated in the 2LAC catalog \citep[][B. Lott personal comm.]{ackermann11}. In Fig.~\ref{mod_nu} we plot the intrinsic modulation indices from the PTF sample against the peak frequencies. The HSP sources are less variable than the LSP sources that have a tail to much higher intrinsic modulation indices.

\section{Correlation between optical and $\gamma$-ray fluxes}\label{sect:flux}
We use the method of \cite{pavlidou12} to study the correlation between PTF and $\gamma$-ray fluxes from the 2FGL catalog \citep{nolan12}. The strength of the correlation is quantified using the Pearson product-moment correlation coefficient $r$ while the significance of the correlation is determined using simulated samples\footnote{The exact method for calculating the correlation coefficient is not critical because by estimating the significance through simulations, we ensure that the results do not depend on the distribution of the variables.}. In order to account for the common redshift in the two variables, the sampling is done in the luminosity space. Because of this, we can only use objects with known redshifts in the correlation analysis. Additionally, we only use simultaneous PTF and 2FGL data and select those PTF sources that had at least 3 data points between 2008-Aug-4 and 2010-Aug-1 (the integration period for 2FGL) to calculate the intrinsic mean flux density. This results in a sample of 118 objects.

We first calculate the rest-frame optical luminosities for all our objects using 
\begin{equation}
L(\nu) = S(\nu)4\pi d^2(1+z)^{1-\alpha},
\end{equation}
where $S(\nu)$ is the PTF intrinsic mean flux density in mJy, calculated using the likelihood method described in the previous section, $d$ is the luminosity distance to the object, $z$ is the redshift and $\alpha$ is the spectral index in the optical band defined as $S(\nu) \propto \nu^\alpha$. We do not have $\alpha$ values available for the individual sources and use values $\alpha=-1.5$ for LSP objects and $\alpha=-1.1$ for the HSP objects based on average values determined in \citet{fiorucci04}. For sources without SED classification or ISP objects we use $\alpha=-1.3$. We note that the exact value of $\alpha$ does not have an effect on the significance of the correlation since we use the same value for both our data and the simulated samples.

The rest-frame $\gamma$-ray luminosities at $E_\gamma = 1$\,GeV are calculated using
\begin{equation}
L(E_\gamma) = (\Gamma-1)F\left(\frac{E_\gamma}{E_0}\right)^{-\Gamma+1}4\pi d^2(1+z)^\Gamma,
\end{equation}
where $F$ is the photon flux above $E_0 = 1$\,GeV and $\Gamma$ is the powerlaw index, both taken from the 2FGL.  According to \citet{nolan12} (see their Fig.~6) the photon fluxes above $E_0 = 1$\,GeV are not affected by the spectral shape if the photon flux is larger than $0.4\times10^{-9}$\,ph\,cm$^{-2}$s$^{-1}$. Our sample includes only four sources below this limit (3 LSP and 1 ISP) and therefore our sample is a nearly unbiased subset of a flux limited $\gamma$-ray sample.

We then construct simulated uncorrelated samples of the same size as our true sample by pairing optical and $\gamma$-ray luminosities of different sources. We move back to the flux-space by assigning a common redshift to the mixed pairs and estimate the strength of the correlation in the intrinsically uncorrelated sample using the Pearson correlation coefficient. Whenever the sample size allows, we use multiple redshift bins in the construction of the simulated samples. In this way we are able to obtain uncorrelated samples that are similar in dynamic range as our real observations. We repeat this procedure $10^7$ times to estimate the significance (p-value) of the correlation. For a more detailed description of the method and recommendations for the redshift binning, see \cite{pavlidou12}.

\begin{figure*}
\includegraphics[scale=0.5]{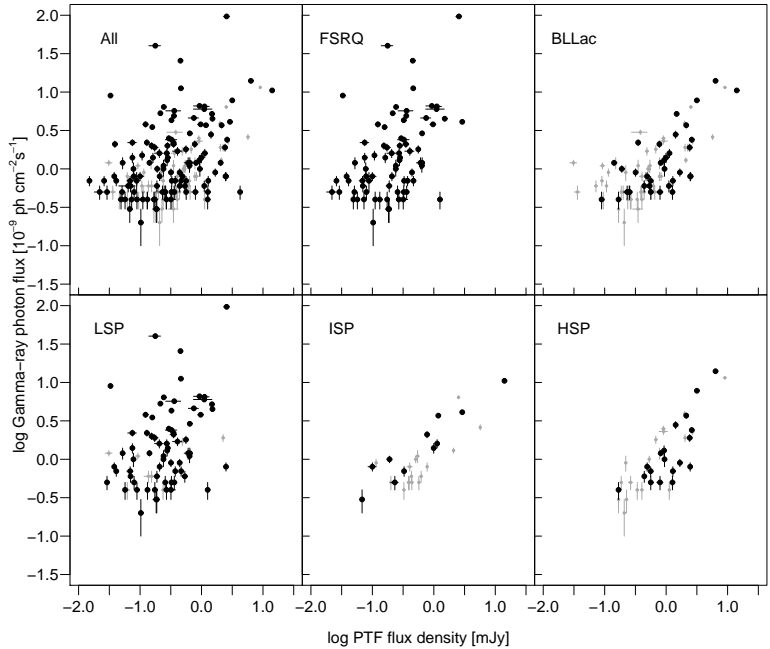}
\caption{Integrated $\gamma$-ray photon flux with its 1$\sigma$ uncertainty plotted against the intrinsic mean optical flux density and its 1$\sigma$ uncertainty for all sources in our sample (top left), FSRQs (top middle), BL~Lac objects (top right), LSPs (bottom left), ISPs (bottom middle), and HSPs (bottom right). Sources without redshifts are plotted in gray symbols and are not included in the calculation of the correlation.}\label{flux}
\end{figure*}

\begin{table}
\centering
\caption{Pearson product-moment correlation coefficient $r$ for the $\gamma$-ray and optical flux correlations and the significance $p$ of the correlations in the various subsamples. Number of sources $N$ in each sample is also listed.}\label{table:flux}
\begin{tabular}{lrrrrrr}
\hline
& $r$ & $p$ & $N$ \\
\hline
All & 0.39 & 0.0043 & 118 \\
FSRQ & 0.46  & 0.0031 & 76 \\
BL~Lac & 0.69 & $5.8\times10^{-5}$ & 34 \\
LSP & 0.36 & 0.0229 & 69 \\
ISP & 0.94 & 0.0001 & 11 \\
HSP & 0.79 & 0.0005 & 19 \\
\hline
\end{tabular}
\end{table} 

Figure~\ref{flux} shows the 2FGL $\gamma$-ray photon flux against the PTF mean optical flux density for the various subclasses. The correlations, summarized in Table~\ref{table:flux}, are at least 3$\sigma$ significant in all subclasses except for the LSP sources and all sources together. There are some caveats in the interpretation of the correlations. Firstly we do not include upper limits in the optical or $\gamma$-ray bands in the correlation estimation. As shown in \cite{lister11}, this can have a significant impact on the result. We may be missing objects both in the upper left (faint in optical) and lower right ($\gamma$-ray faint) corners of the plot which will affect the strength  and the scatter of the correlation, although not its significance.
Secondly, we have not accounted for the host-galaxy emission in calculation of the optical fluxes, which may have an effect on the fluxes of the low redshift objects \citep[e.g.,][]{urry00,falomo03}. Furthermore, the optical emission in FSRQs can have a significant thermal component \citep[e.g.,][]{raiteri07} which may increase the scatter in any correlations.

\section{Discussion}\label{sect:discussion}
Using a large sample of sources from the PTF and CRTS optical surveys, we have studied the variability properties of $\gamma$-ray detected and non-detected objects, and
of BL~Lacs and FSRQs, in addition to dividing the sources based on their SED classification. We have determined the {\it intrinsic modulation index} for the different source populations and studied the flux - flux correlation between the optical and $\gamma$ rays.

There are 637 sources in common in the PTF and CRTS samples. These include 320 sources with detected variability at more than $3\sigma$ level. A notable result seen in Figs.~\ref{gammadet}-\ref{sedclass} is that the mean intrinsic modulation indices of the CRTS sources are systematically higher than in PTF. The number of data points in the two surveys are comparable but the CRTS data are more uniformly distributed over the five-year period considered here. This indicates that in order to detect the largest variability amplitudes long-term monitoring is needed. However, all our main findings are seen in both PTF and CRTS data sets, confirming the intrinsic nature of the differences between the sub populations.

\begin{figure}
\includegraphics[scale=0.6]{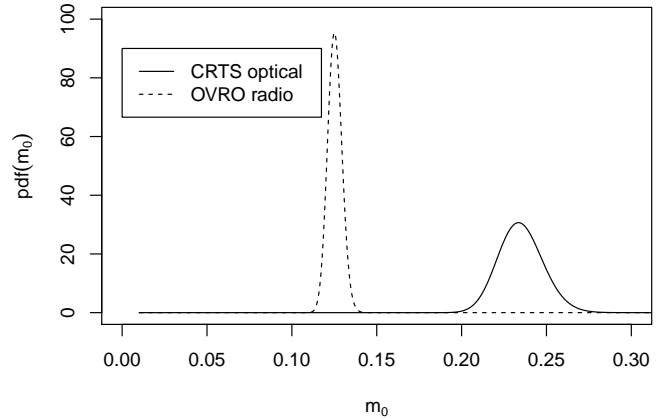}
\caption{Probability density of the mean intrinsic modulation index for OVRO and CRTS data. Solid line is CRTS data (maximum-likelihood value and its 1$\sigma$ error $m_0 = 0.234^{+0.014}_{-0.013}$ ). Dashed line is OVRO data ($m_0 = 0.125^{+0.004}_{-0.004}$).}\label{crts_ovro}
\end{figure}
Our sample was selected from the monitoring sample at OVRO so as to allow for comparison between the optical and 15\,GHz radio variability. We use the CRTS sample for the comparison because it is larger (1244 objects) and the more uniform sampling is more similar to the OVRO observing cadence (on average 2 observations per week). The OVRO modulation indices are calculated based on four years of data \citep{richards14}. Fig.~\ref{crts_ovro} shows the probability density of the mean modulation index for the overlapping CRTS and OVRO samples. The variability in the optical ($m_0 = 0.234^{+0.014}_{-0.013}$) is nearly twice as high as in the radio ($m_0 = 0.125^{+0.004}_{0.004}$). This is expected because the optical emission is thought to originate in smaller emission regions and be caused by higher energy electrons near the peak of the electron distribution.

\subsection{Optical variability}
One of our main results is that the {\it Fermi}-detected sources are more variable in optical than the non-detected objects. This has important implications for the 
identification of the unassociated objects in the {\it Fermi} catalogs. Up to 30\% of the sources in the 2FGL catalog lack associations at other wavebands
\citep{nolan12}. This is mainly because the source position errors are still fairly large and the 95\% error circle can include multiple objects. In order to 
identify the correct counterpart for the $\gamma$-ray source some additional information is needed. Identifying unassociated sources has successfully been done by correlating the $\gamma$-ray 
properties with known source populations \citep{ackermann12b} and by using data at various other wavelengths \citep[e.g.,][]{kovalev09b,dabrusco13,massaro13b,massaro13,maselli13}.

\cite{ruan12} showed how the variability of sources in optical catalogs can be used to identify the blazar counterparts of the {\it Fermi} objects. They used optical 
data from the LINEAR asteroid survey and characterized the variability timescales and amplitudes of the objects. By selecting sources with variability characteristics similar to those of known blazars, they were able to recover 88\% of the known associations in the 2FGL catalog, showing that identifying blazars using optical variability is 
an efficient tool. Our results agree with this conclusion and we suggest that the PTF and CRTS data can be used to aid in the identification of the unassociated sources (A. Mahabal et al. in prep.). One advantage of our modulation index method is that it can be used even for sources with only a few data points, as was explained in Sect.~\ref{sect:modindex}. 

Our result agrees with the analysis of the OVRO 15\,GHz radio data by \cite{richards11} that found the $\gamma$-ray detected objects to be more variable than 
the non-detected ones. They suggest that one possibility is that all the radio-loud objects are also $\gamma$-ray loud but were not detected by the LAT because they 
were not flaring during the first year of {\it Fermi} operations. This hypothesis was also suggested by \cite{kovalev09} who showed that the sources detected during the first three months of LAT monitoring were more variable in the 15\,GHz Very Long Baseline Array data than the non-detected ones. 

In addition, we divided our sources based on their optical classification into BL~Lacs and FSRQs and based on their SEDs into LSP, ISP and HSP sources. When looking at the total sample we find the FSRQs to be significantly more variable than BL~Lacs. This is mainly due to the location of the SED peak so that the lower variability in BL~Lacs is driven by the less variable ISP and HSP sources. In fact, if we compare  the CGRaBS sources only (dominated by LSP BL~Lacs), or the {\it Fermi} -detected LSP BL~Lacs and FSRQs, the difference between the two optical classes is significant only at the 1$\sigma$ level. Similar results were obtained by \cite{ikejiri11}  who studied the NIR variability in a sample of 42 objects. This could potentially be due to many of the LSP BL~Lacs being intrinsically similar to FSRQs as suggested by \cite{giommi12,giommi13}, but falsely classified as BL~Lacs due to the jet emission swamping any thermal emission.

A similar tendency for higher variability in the LSP sources was found in the $\gamma$-ray data in the 2nd LAT AGN catalog \citep{ackermann11}. They attribute this to the location of the high-energy peak with respect to the {\it Fermi} band. LSP sources observed by {\it Fermi} are observed at greater energies than the inverse Compton peak and therefore are produced by higher energy electrons which cool faster and vary more. HSP sources on the other hand are observed at lower energies than the peak and therefore would vary less. This is similar to what we see in the optical observations with respect to the synchrotron peak \citep[e.g.,][]{mastichiadis97,kirk99}.  Interestingly, \cite{richards11} found using 15 GHz data that BL~Lac objects were more variable than FSRQs in the CGRaBS sample. This can at least partially be explained by the CGRaBS BL~Lac sample being dominated by LSP objects which tend to vary more than the HSPs. When a gamma-ray--selected parent sample is used instead of CGRaBS, no significant difference in variability is found between FSRQs and BL~Lacs, and a trend with greater variability in LSPs than HSPs is seen \citep{richards14}.

\cite{bauer09} used the Palomar QUEST survey to extract optical variability information for blazars and also found that FSRQs exhibit larger amplitude variations than the BL~Lac objects. They suggest that this could be due to higher jet power in the FSRQs which would allow them to produce larger flares. This agrees with the suggestion of \cite{ghisellini98} that more luminous objects have higher energy density and therefore cool faster which causes their synchrotron spectrum to peak at lower frequencies. \cite{ghisellini98} used this to explain the so called blazar sequence \citep{fossati98} where the peak of the synchrotron and inverse Compton components shifts in frequency and luminosity depending on the source type. Subsequently, several issues have been identified with the original blazar sequence \citep[e.g.,][]{padovani07,nieppola08} and \cite{ghisellini08} updated their model to include the accretion disc differences between the objects. They suggest that the high power sources have an efficiently radiating accretion disc that contributes to their emission while the lower power BL~Lac objects have an inefficient accretion disc. Recently \cite{meyer11} showed using larger samples that, instead of a simple sequence the FSRQs and BL~Lacs have different loci in the jet power - synchrotron peak plane which can explain some of the differences in their observed properties. We may also be seeing an additional contribution from the accretion disc in the FSRQs in the optical band. Combining the effects of higher jet power, more efficient accretion disc, and higher electron energies at the optical band in FSRQs compared to BL~Lacs can most likely explain the difference in variability we see in these objects. 

\subsection{Flux correlations}
In addition to finding that the activity in the two bands is connected, we also find the simultaneous fluxes to be correlated. We find the correlation to have less scatter in BL~Lacs than the FSRQs. This is opposite to what was seen by \cite{arshakian12} who used non-simultaneous data to study the optical - $\gamma$-ray flux correlation in 82 sources and found a significant correlation only in the FSRQs. We found that when testing for the correlation using the entire PTF data set without requiring simultaneous data, the scatter in the correlation was much larger. As suggested by \cite{arshakian12} use of non-simultaneous data can most likely explain the large scatter seen in their correlations.

If the $\gamma$-ray emission is produced by the synchrotron self-Compton (SSC) method, we would expect a tight correlation in the flux-flux space due to the same amount of Doppler-boosting in these regimes. In the case of external Compton emission, the high-energy component is more boosted which destroys a linear dependence \citep[e.g.,][]{dermer95}. Additionally, there can be a large contribution from the external thermal photon field to the observed optical fluxes, hindering any correlations further. It is typically possible to model the SEDs of HSP sources using just a single SSC component while the FSRQs and LSP sources nearly always require an additional EC component \citep[e.g.,][]{lindfors05,abdo10c,abdo11,bottcher13}.

Our observations are in good agreement with these models with the ISP and HSP sources showing a tighter correlation than the FSRQs. A Similar conclusion was drawn by \cite{lister11} based on a tight correlation between the $\gamma$-ray to radio flux ratio and the synchrotron peak frequency in a complete radio and $\gamma$-ray selected sample of 173 AGNs. One caveat in the comparison using radio data is that the radio and $\gamma$-ray emission are produced by very different energy electrons. 
The higher $\gamma$-ray to radio ratios in HSP objects in comparison to LSP sources seen in \cite{lister11} and in \cite{nieppola11} can be explained by the shifting of the synchrotron peak to higher frequencies in HSP objects, which decreases the amount of radio emission seen and increases the $\gamma$-ray dominance. 

\begin{figure}
\includegraphics[scale=0.4]{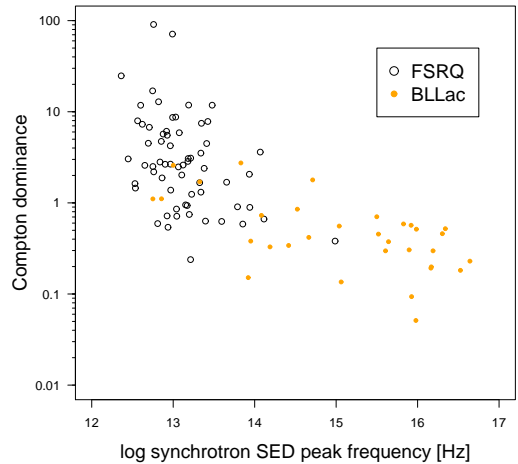}
\caption{Compton dominance (ratio between $\gamma$-ray and PTF $\nu F_\nu$) against the synchrotron peak frequency. Open circles are FSRQs and orange filled circles BL~Lac objects.}\label{compton}
\end{figure}

Alternatively, the ratio between $\gamma$-ray and optical emission is nearly the same as the Compton dominance, defined as the ratio of the peak luminosity in the IC and synchrotron components. As in \cite{pavlidou12}, we use the photon flux between 1 and 100\,GeV tabulated in 2FGL to calculate the $\gamma$-ray flux density as 
\begin{equation}
S_E(E_\gamma) = (\Gamma-1)F\left(\frac{E_\gamma}{E_0}\right)^{(1-\Gamma)} \rm{cm}^{-2}\rm{s}^{-1},
\end{equation}
where $\Gamma$ is the powerlaw index, $F$ is the photon flux in units ph~cm$^{-2}$~s$^{-1}$, $E_\gamma$ is the energy where we want to define the energy flux density (in our case 1~GeV) and $E_0$ is the lower limit in energy (1 GeV in our case). Strictly, this equation assumes the photon flux is integrated from $E_0$ to infinity, but due to the steep photon indices, the difference to limiting the equation at 100~GeV is less than 1\%. Expressing this in the SED units in the rest frame of the source, we find 
\begin{equation}
E_\gamma\cdot S(E_\gamma) = E_\gamma \cdot S_E(E_\gamma)(1+z)^\Gamma,
\end{equation}
where $E_\gamma$ is the energy at which to estimate the flux in erg (in our case $1.6\times10^{-3}$ erg corresponding to 1 GeV).
Similarly, we convert our simultaneous PTF mean intrinsic flux densities to rest frame SED units at the central frequency $\nu = 4.56\times10^{14}$~Hz (658~nm), which gives
\begin{equation}
\nu F_\nu = \nu S(\nu)(1+z)^{(1-\alpha)},
\end{equation}
where $S(\nu)$ is the flux density in erg~cm$^{-2}$~s$^{-1}$~Hz$^{-1}$, and $\alpha$ is the spectral index in the optical band. We then
calculate the ratio ($E_\gamma\cdot S(E_\gamma))/(\nu F(\nu))$ which we call the Compton dominance. 

In Fig.~\ref{compton} we plot the Compton dominance against the synchrotron peak frequency. The LSPs are much more Compton dominated than the HSPs as was also seen by \cite{abdo10c}. They attribute this to the higher EC contribution in the LSP objects. This agrees very well with our flux-flux correlation result. Our results are also very similar to \cite{finke13} who studied the Compton dominance in {\it Fermi} blazars by estimating the synchrotron and IC peak SED luminosities. \cite{finke13} explain the trends seen in the plot with a simple model where the difference between the sources is due to magnetic field and energy density in the emission region, in addition to the viewing angle of the source. 

Taking the connected variations in the two bands together with the flux-flux correlation is a strong indication for SSC origin of $\gamma$-ray emission in the HSP objects. However, we note that in individual sources and in individual flares the situation may be more complicated as shown in several detailed studies of individual sources \citep[e.g.,][]{marscher10,ackermann12,nalewajko12,orienti13}. 

\section{Conclusions}\label{sect:conclusions}
We have studied the optical variability of AGN using large samples of objects from the PTF and CRTS surveys. We use a likelihood approach to calculate the intrinsic modulation index and a maximum likelihood method to study the differences between various source populations. Additionally, we studied the flux-flux correlation between the optical and $\gamma$-ray bands. Our main results can be summarized as follows:
\begin{enumerate}
\item We find the {\it Fermi}-detected objects to be more variable than non-detected ones. This shows that the activity in the two bands is connected and that the optical variability can be used as a tool for identifying unassociated {\it Fermi} sources.
\item The FSRQs in our  total sample are more variable than the BL~Lac objects. This is likely due to the location of their synchrotron peak because the mean modulation indices of  radio-selected CGRaBS FSRQs and BL~Lacs or the {\it Fermi}-detected FSRQs and LSP BL~Lacs do not differ significantly.
\item When dividing the objects based on their synchrotron peak location, we find the LSP objects to be more variable than the HSPs, with ISPs in between the two. This is similar to what is seen in the $\gamma$-ray band and can be due to differences in the electron energy in the observed bands. 
\item We find a significant correlation between the optical and $\gamma$-ray fluxes which is tighter in the BL~Lac objects. The FSRQs are also more Compton dominated than the BL~Lacs. This is in accordance with models where the high-energy emission of the HSP and ISP sources can be modeled with a single SSC component while in FSRQs an additional EC component is required.
\end{enumerate}

\section*{Acknowledgments}
We thank the referee, Tigran Arshakian, for useful comments that helped to improve the manuscript. T.H. thanks Elina Lindfors for useful discussions.
T.H. was supported in part by the Jenny and Antti Wihuri foundation. V.P. is acknowledging support by the European Comission Seventh Framework Programme (FP7) through the Marie Curie Career Integration Grant PCIG10-GA-2011-304001 ``JetPop'' and the EU FP7 Grant PIRSES-GA-2012-31578 ``EuroCal''. The OVRO 40-m monitoring program is supported in part by NASA grants NNX08AW31G and NNX11A043G, and NSF grants AST-0808050 and AST-1109911. The National Radio Astronomy Observatory is a facility of the National Science Foundation operated under cooperative agreement by Associated Universities Inc.
The CSS survey is funded by the National Aeronautics and Space
Administration under Grant No. NNG05GF22G issued through the Science
Mission Directorate Near-Earth Objects Observations Program.  The CRTS
survey is supported by the U.S.~National Science Foundation under
grants AST-0909182.
This article is based on observations obtained with the Samuel Oschin Telescope as part of the Palomar Transient Factory project, a scientific collaboration between the California Institute of Technology, Columbia University, Las Cumbres Observatory, the Lawrence Berkeley National Laboratory, the National Energy Research Scientific Computing Center, the University of Oxford, and the Weizmann Institute of Science. It is also partially based on observations obtained as part of the Intermediate Palomar Transient Factory project, a scientific collaboration among the California Institute of Technology, Los Alamos National Laboratory, the University of Wisconsin, Millwakee, the Oskar Klein Center, the Weizmann Institute of Science, the TANGO Program of the University System of Taiwan, the Kavli Institute for the Physics and Mathematics of the Universe, and the Inter-University Centre for Astronomy and Astrophysics.
This research has made use of the NASA/IPAC Extragalactic Database (NED) which is operated by the Jet Propulsion Laboratory, California Institute of Technology, under contract with the National Aeronautics and Space Administration. 

\footnotesize{
\bibliographystyle{mn2eb}
% Use the LaTeX power, use bibtex properly.                                                                                                                              
\bibliography{thbib}

\begin{thebibliography}{79}
\providecommand{\natexlab}[1]{#1}

\bibitem[{{Abdo} et~al.(2010{\natexlab{a}})}]{abdo10}
{Abdo} A.~A. et~al., 2010{\natexlab{a}}, \nat, 463, 919

\bibitem[{{Abdo} et~al.(2010{\natexlab{b}})}]{abdo10b}
{Abdo} A.~A. et~al., 2010{\natexlab{b}}, \apj, 715, 429

\bibitem[{{Abdo} et~al.(2010{\natexlab{c}})}]{abdo10c}
{Abdo} A.~A. et~al., 2010{\natexlab{c}}, \apj, 716, 30

\bibitem[{{Abdo} et~al.(2011)}]{abdo11}
{Abdo} A.~A. et~al., 2011, \apj, 736, 131

\bibitem[{{Ackermann} et~al.(2011)}]{ackermann11}
{Ackermann} M. et~al., 2011, \apj, 743, 171

\bibitem[{{Ackermann} et~al.(2012{\natexlab{a}})}]{ackermann12b}
{Ackermann} M. et~al., 2012{\natexlab{a}}, \apj, 753, 83

\bibitem[{{Ackermann} et~al.(2012{\natexlab{b}})}]{ackermann12}
{Ackermann} M. et~al., 2012{\natexlab{b}}, \apj, 751, 159

\bibitem[{{Agudo} et~al.(2011)}]{agudo11}
{Agudo} I. et~al., 2011, \apjl, 726, L13

\bibitem[{{Aller} et~al.(1992){Aller}, {Aller} \& {Hughes}}]{aller92}
{Aller} M.~F., {Aller} H.~D., {Hughes} P.~A., 1992, \apj, 399, 16

\bibitem[{{Arshakian} et~al.(2012)}]{arshakian12}
{Arshakian} T.~G., {Le{\'o}n-Tavares} J., {B{\"o}ttcher} M., {Torrealba} J.,
  {Chavushyan} V.~H., {Lister} M.~L., {Ros} E., {Zensus} J.~A., 2012, \aap,
  537, A32

\bibitem[{{Atwood} et~al.(2009)}]{atwood09}
{Atwood} W.~B. et~al., 2009, \apj, 697, 1071

\bibitem[{{Bauer} et~al.(2009){Bauer}, {Baltay}, {Coppi}, {Ellman}, {Jerke},
  {Rabinowitz} \& {Scalzo}}]{bauer09}
{Bauer} A., {Baltay} C., {Coppi} P., {Ellman} N., {Jerke} J., {Rabinowitz} D.,
  {Scalzo} R., 2009, \apj, 699, 1732

\bibitem[{{Bertin} \& {Arnouts}(1996)}]{bertin96}
{Bertin} E., {Arnouts} S., 1996, \aaps, 117, 393

\bibitem[{{Bloom} \& {Marscher}(1996)}]{bloom96}
{Bloom} S.~D., {Marscher} A.~P., 1996, \apj, 461, 657

\bibitem[{{Bloom} et~al.(1997)}]{bloom97}
{Bloom} S.~D. et~al., 1997, \apjl, 490, L145

\bibitem[{{Bonning} et~al.(2012)}]{bonning12}
{Bonning} E. et~al., 2012, \apj, 756, 13

\bibitem[{{Bonning} et~al.(2009)}]{bonning09}
{Bonning} E.~W. et~al., 2009, \apjl, 697, L81

\bibitem[{{B{\"o}ttcher} et~al.(2013){B{\"o}ttcher}, {Reimer}, {Sweeney} \&
  {Prakash}}]{bottcher13}
{B{\"o}ttcher} M., {Reimer} A., {Sweeney} K., {Prakash} A., 2013, \apj, 768, 54

\bibitem[{{Chatterjee} et~al.(2012)}]{chatterjee12}
{Chatterjee} R. et~al., 2012, \apj, 749, 191

\bibitem[{{D'Abrusco} et~al.(2013){D'Abrusco}, {Massaro}, {Paggi}, {Masetti},
  {Tosti}, {Giroletti} \& {Smith}}]{dabrusco13}
{D'Abrusco} R., {Massaro} F., {Paggi} A., {Masetti} N., {Tosti} G., {Giroletti}
  M., {Smith} H.~A., 2013, \apjs, 206, 12

\bibitem[{{Dermer}(1995)}]{dermer95}
{Dermer} C.~D., 1995, \apjl, 446, L63

\bibitem[{{Dermer} \& {Schlickeiser}(1993)}]{dermer93}
{Dermer} C.~D., {Schlickeiser} R., 1993, \apj, 416, 458

\bibitem[{{Djorgovski} et~al.(2012)}]{djorgovski12}
{Djorgovski} S.~G. et~al., 2012, in T.~{Mihara}, M.~{Serino}, eds, The First
  Year of MAXI: Monitoring Variable X-ray Sources. p. 263

\bibitem[{{Drake} et~al.(2009)}]{drake09}
{Drake} A.~J. et~al., 2009, \apj, 696, 870

\bibitem[{{Edelson} et~al.(2002)}]{edelson02}
{Edelson} R., {Turner} T.~J., {Pounds} K., {Vaughan} S., {Markowitz} A.,
  {Marshall} H., {Dobbie} P., {Warwick} R., 2002, \apj, 568, 610

\bibitem[{{Falomo} et~al.(2003){Falomo}, {Carangelo} \& {Treves}}]{falomo03}
{Falomo} R., {Carangelo} N., {Treves} A., 2003, \mnras, 343, 505

\bibitem[{{Fanaroff} \& {Riley}(1974)}]{fanaroff74}
{Fanaroff} B.~L., {Riley} J.~M., 1974, \mnras, 167, 31P

\bibitem[{{Finke}(2013)}]{finke13}
{Finke} J.~D., 2013, \apj, 763, 134

\bibitem[{{Fiorucci} et~al.(2004){Fiorucci}, {Ciprini} \& {Tosti}}]{fiorucci04}
{Fiorucci} M., {Ciprini} S., {Tosti} G., 2004, \aap, 419, 25

\bibitem[{{Fitzpatrick}(1999)}]{fitzpatrick99}
{Fitzpatrick} E.~L., 1999, \pasp, 111, 63

\bibitem[{{Fossati} et~al.(1998){Fossati}, {Maraschi}, {Celotti}, {Comastri} \&
  {Ghisellini}}]{fossati98}
{Fossati} G., {Maraschi} L., {Celotti} A., {Comastri} A., {Ghisellini} G.,
  1998, \mnras, 299, 433

\bibitem[{{Ghisellini} \& {Tavecchio}(2008)}]{ghisellini08}
{Ghisellini} G., {Tavecchio} F., 2008, \mnras, 387, 1669

\bibitem[{{Ghisellini} et~al.(1998){Ghisellini}, {Celotti}, {Fossati},
  {Maraschi} \& {Comastri}}]{ghisellini98}
{Ghisellini} G., {Celotti} A., {Fossati} G., {Maraschi} L., {Comastri} A.,
  1998, \mnras, 301, 451

\bibitem[{{Giommi} et~al.(2012){Giommi}, {Padovani}, {Polenta}, {Turriziani},
  {D'Elia} \& {Piranomonte}}]{giommi12}
{Giommi} P., {Padovani} P., {Polenta} G., {Turriziani} S., {D'Elia} V.,
  {Piranomonte} S., 2012, \mnras, 420, 2899

\bibitem[{{Giommi} et~al.(2013){Giommi}, {Padovani} \& {Polenta}}]{giommi13}
{Giommi} P., {Padovani} P., {Polenta} G., 2013, \mnras, 431, 1914

\bibitem[{{Hartman} et~al.(2001)}]{hartman01}
{Hartman} R.~C. et~al., 2001, \apj, 558, 583

\bibitem[{{Healey} et~al.(2008)}]{healey08}
{Healey} S.~E. et~al., 2008, \apjs, 175, 97

\bibitem[{{Ikejiri} et~al.(2011)}]{ikejiri11}
{Ikejiri} Y. et~al., 2011, \pasj, 63, 639

\bibitem[{{Jester} et~al.(2005)}]{jester05}
{Jester} S. et~al., 2005, \aj, 130, 873

\bibitem[{{Kharb} et~al.(2010){Kharb}, {Lister} \& {Cooper}}]{kharb10}
{Kharb} P., {Lister} M.~L., {Cooper} N.~J., 2010, \apj, 710, 764

\bibitem[{{Kirk} \& {Mastichiadis}(1999)}]{kirk99}
{Kirk} J.~G., {Mastichiadis} A., 1999, Astroparticle Physics, 11, 45

\bibitem[{{Kovalev}(2009)}]{kovalev09b}
{Kovalev} Y.~Y., 2009, \apjl, 707, L56

\bibitem[{{Kovalev} et~al.(2009)}]{kovalev09}
{Kovalev} Y.~Y. et~al., 2009, \apjl, 696, L17

\bibitem[{{Kraus} et~al.(2003)}]{kraus03}
{Kraus} A. et~al., 2003, \aap, 401, 161

\bibitem[{{Landoni} et~al.(2012){Landoni}, {Falomo}, {Treves}, {Sbarufatti},
  {Decarli}, {Tavecchio} \& {Kotilainen}}]{landoni12}
{Landoni} M., {Falomo} R., {Treves} A., {Sbarufatti} B., {Decarli} R.,
  {Tavecchio} F., {Kotilainen} J., 2012, \aap, 543, A116

\bibitem[{{Landt} \& {Bignall}(2008)}]{landt08}
{Landt} H., {Bignall} H.~E., 2008, \mnras, 391, 967

\bibitem[{{Law} et~al.(2009)}]{law09}
{Law} N.~M. et~al., 2009, \pasp, 121, 1395

\bibitem[{{Lindfors} et~al.(2005){Lindfors}, Valtaoja \& M.}]{lindfors05}
{Lindfors} E.~J., Valtaoja E., M. T., 2005, \aap, 440, 845

\bibitem[{{Lister} et~al.(2011)}]{lister11}
{Lister} M.~L. et~al., 2011, \apj, 742, 27

\bibitem[{{Mahabal} et~al.(2011)}]{mahabal11}
{Mahabal} A.~A. et~al., 2011, Bulletin of the Astronomical Society of India,
  39, 387

\bibitem[{{Maraschi} et~al.(1992){Maraschi}, {Ghisellini} \&
  {Celotti}}]{maraschi92}
{Maraschi} L., {Ghisellini} G., {Celotti} A., 1992, \apjl, 397, L5

\bibitem[{{Marscher} et~al.(2010)}]{marscher10}
{Marscher} A.~P. et~al., 2010, \apjl, 710, L126

\bibitem[{{Maselli} et~al.(2013)}]{maselli13}
{Maselli} A. et~al., 2013, \apjs, 206, 17

\bibitem[{{Massaro} et~al.(2013{\natexlab{a}})}]{massaro13b}
{Massaro} F., {D'Abrusco} R., {Giroletti} M., {Paggi} A., {Masetti} N., {Tosti}
  G., {Nori} M., {Funk} S., 2013{\natexlab{a}}, \apjs, 207, 4

\bibitem[{{Massaro} et~al.(2013{\natexlab{b}})}]{massaro13}
{Massaro} F., {D'Abrusco} R., {Paggi} A., {Masetti} N., {Giroletti} M., {Tosti}
  G., {Smith} H.~A., {Funk} S., 2013{\natexlab{b}}, \apjs, 206, 13

\bibitem[{{Mastichiadis} \& {Kirk}(1997)}]{mastichiadis97}
{Mastichiadis} A., {Kirk} J.~G., 1997, \aap, 320, 19

\bibitem[{{Meyer} et~al.(2011){Meyer}, {Fossati}, {Georganopoulos} \&
  {Lister}}]{meyer11}
{Meyer} E.~T., {Fossati} G., {Georganopoulos} M., {Lister} M.~L., 2011, \apj,
  740, 98

\bibitem[{{Nalewajko} et~al.(2012)}]{nalewajko12}
{Nalewajko} K., {Sikora} M., {Madejski} G.~M., {Exter} K., {Szostek} A.,
  {Szczerba} R., {Kidger} M.~R., {Lorente} R., 2012, \apj, 760, 69

\bibitem[{{Nieppola} et~al.(2008){Nieppola}, {Valtaoja}, {Tornikoski},
  {Hovatta} \& {Kotiranta}}]{nieppola08}
{Nieppola} E., {Valtaoja} E., {Tornikoski} M., {Hovatta} T., {Kotiranta} M.,
  2008, \aap, 488, 867

\bibitem[{{Nieppola} et~al.(2011){Nieppola}, {Tornikoski}, {Valtaoja},
  {Le{\'o}n-Tavares}, {Hovatta}, {L{\"a}hteenm{\"a}ki} \& {Tammi}}]{nieppola11}
{Nieppola} E., {Tornikoski} M., {Valtaoja} E., {Le{\'o}n-Tavares} J., {Hovatta}
  T., {L{\"a}hteenm{\"a}ki} A., {Tammi} J., 2011, \aap, 535, A69

\bibitem[{{Nolan} et~al.(2012)}]{nolan12}
{Nolan} P.~L. et~al., 2012, \apjs, 199, 31

\bibitem[{{Ofek} et~al.(2011)}]{ofek11}
{Ofek} E.~O., {Frail} D.~A., {Breslauer} B., {Kulkarni} S.~R., {Chandra} P.,
  {Gal-Yam} A., {Kasliwal} M.~M., {Gehrels} N., 2011, \apj, 740, 65

\bibitem[{{Ofek} et~al.(2012{\natexlab{a}})}]{ofek12a}
{Ofek} E.~O. et~al., 2012{\natexlab{a}}, \pasp, 124, 62

\bibitem[{{Ofek} et~al.(2012{\natexlab{b}})}]{ofek12b}
{Ofek} E.~O. et~al., 2012{\natexlab{b}}, \pasp, 124, 854

\bibitem[{{Orienti} et~al.(2013)}]{orienti13}
{Orienti} M. et~al., 2013, \mnras, 428, 2418

\bibitem[{{Padovani}(2007)}]{padovani07}
{Padovani} P., 2007, \apss, 309, 63

\bibitem[{{Pavlidou} et~al.(2012)}]{pavlidou12}
{Pavlidou} V. et~al., 2012, \apj, 751, 149

\bibitem[{{Raiteri} et~al.(2007)}]{raiteri07}
{Raiteri} C.~M. et~al., 2007, \aap, 473, 819

\bibitem[{{Rau} et~al.(2009)}]{rau09}
{Rau} A. et~al., 2009, \pasp, 121, 1334

\bibitem[{{Richards} et~al.(2014){Richards}, Hovatta, Max-Moerbeck, {Pavlidou},
  Pearson \& Readhead}]{richards14}
{Richards} J.~L., Hovatta T., Max-Moerbeck W., {Pavlidou} V., Pearson T.~J.,
  Readhead A.~C.~S., 2014, MNRAS in press

\bibitem[{{Richards} et~al.(2011)}]{richards11}
{Richards} J.~L. et~al., 2011, \apjs, 194, 29

\bibitem[{{Ruan} et~al.(2012)}]{ruan12}
{Ruan} J.~J. et~al., 2012, \apj, 760, 51

\bibitem[{{Schlafly} \& {Finkbeiner}(2011)}]{schlafly11}
{Schlafly} E.~F., {Finkbeiner} D.~P., 2011, \apj, 737, 103

\bibitem[{{Shaw} et~al.(2013)}]{shaw13}
{Shaw} M.~S. et~al., 2013, \apj, 764, 135

\bibitem[{{Sikora} et~al.(1994){Sikora}, {Begelman} \& {Rees}}]{sikora94}
{Sikora} M., {Begelman} M.~C., {Rees} M.~J., 1994, \apj, 421, 153

\bibitem[{{Urry} et~al.(2000){Urry}, {Scarpa}, {O'Dowd}, {Falomo}, {Pesce} \&
  {Treves}}]{urry00}
{Urry} C.~M., {Scarpa} R., {O'Dowd} M., {Falomo} R., {Pesce} J.~E., {Treves}
  A., 2000, \apj, 532, 816

\bibitem[{{Wagner} \& {Witzel}(1995)}]{wagner95}
{Wagner} S.~J., {Witzel} A., 1995, \araa, 33, 163

\bibitem[{{Wagner} et~al.(1995{\natexlab{a}})}]{wagner95c}
{Wagner} S.~J. et~al., 1995{\natexlab{a}}, \apjl, 454, L97

\bibitem[{{Wagner} et~al.(1995{\natexlab{b}})}]{wagner95b}
{Wagner} S.~J. et~al., 1995{\natexlab{b}}, \aap, 298, 688

\end{thebibliography}
}
\label{lastpage}

\begin{landscape}
\begin{table}
\begin{minipage}{250mm}
\caption{Sources in the PTF sample. The full table will be available in the online edition of the journal.}\label{table:PTF}
\footnotesize
\begin{tabular}{llrlrcccrr}
\hline
OVRO name  &    Right Ascension  & Declination  &   {\it Fermi}
association   &    Redshift     &   Redshift reference   &
Opt. Class  &    SED class   &    $\overline{m}$ & $S_0$ \\
\hline
CRJ0001$-$0746  &  00:01:18.03 &    $-$07:46:27.01  &  2FGLJ0000.9$-$0748     &   $...$ &   $...$  &  B    &   ISP  &   $0.089_{-0.017}^{+0.023}$ &   $0.446_{-0.011}^{+0.011}$ \\
J0003+2129    &  00:03:19.35   &  21:29:44.40    & $...$  &  0.45  &  1   &    A    &   $...$ &   $<0.146$ &   $0.037_{-0.000}^{+0.000}$ \\
J0005+0524    &  00:05:20.21   &  05:24:10.70    & $...$   & 1.9   &  1    &  Q     &  $...$   & $<0.014$ &     $1.234_{-0.000}^{+0.000}$ \\
J0006$-$0623    &  00:06:13.89   &  -06:23:35.30   &$...$   & 0.347  & 1     &  B    &   $...$   & $0.266_{-0.050}^{+0.072}$ &   $0.225_{-0.017}^{+0.017}$\\
J0006+2422    &  00:06:48.79   &  24:22:36.50    & $...$   & 1.684  & 1    &   Q   &    $...$   & $<0.112$ &  $0.108_{-0.000}^{+0.000}$ \\
J0010+2047    &  00:10:28.74   &  20:47:49.70    & $...$   & 0.6   &  1   &    Q    &   $...$   & $0.162_{-0.024}^{+0.029}$ & $0.085_{-0.003}^{+0.003}$\\
J0010+1058    &  00:10:31.01   &  10:58:29.50    & $...$   & 0.089 &  1   &    A   &    $...$  &  $0.094_{-0.019}^{+0.027}$ &   $4.015_{-0.118}^{+0.117}$\\
J0010+1724    &  00:10:33.99   &  17:24:18.80    & $...$   & 1.601 &  1   &    Q   &    $...$   & $0.090_{-0.011}^{+0.014}$ & $0.646_{-0.011}^{+0.011}$\\
J0013$-$1513    &  00:13:20.71   &  $-$15:13:47.90   & $...$   & 1.838  & 1    &   Q    &   $...$  &  $<0.119$   & $0.050_{-0.000}^{+0.000}$ \\
J0013+1910     & 00:13:56.38   &  19:10:41.90    & 2FGLJ0013.8+1907  &      $...$  &  $...$  &  B  &     $...$  & $0.392_{-0.038}^{+0.045}$ &   $0.185_{-0.009}^{+0.009}$\\
\hline
\end{tabular}
\end{minipage}
%\medskip
Columns are as follows: (1) OVRO Name; (2) RA; (3) DEC; (4) {\it
  Fermi} association; (5) redshift; (6) redshift reference where 1 is
\cite{healey08}, 2 is \cite{abdo10}, 3 is \cite{ackermann11}, and 4 is \cite{landoni12}; (7) optical classification where Q = FSRQ, B = BL~Lac
object, G = galaxy, A = Active galactic nucleus, N = Narrow line
Seyfert I galaxy, and U = unidentified; (8) SED class from Ackermann et
al. (2011); (9) Intrinsic modulation index and its 1$\sigma$ error; (10) Intrinsic mean flux density and its $1\sigma$ error.
%\end{minipage}
\end{table}
\end{landscape}

\begin{landscape}
\begin{table}
\begin{minipage}{250mm}
\caption{Sources in the CRTS sample. The full table will be available in the online edition of the journal.}\label{table:CRTS}
\footnotesize
\begin{tabular}{llrlrcccrr}
\hline
OVRO name  &    Right Ascension  & Declination  &   {\it Fermi}
association   &    Redshift     &   Redshift reference   &
Opt. Class  &    SED class   &    $\overline{m}$ & $S_0$ \\
\hline
J0001$-$1551   &   00:01:5.33  &    $-$15:51:7.10 &    $...$  &  2.044 &  1    &   Q   &    $...$  & $< 0.130$   & $0.215_{-0.000}^{+0.000}$ \\
CRJ0001$-$0746 &   00:01:18.03 &    $-$07:46:27.01  &  2FGLJ0000.9$-$0748    &    $...$  &  $...$  &  B    &   ISP &   $0.327_{-0.044}^{+0.057}$ &   $0.458_{-0.028}^{+0.028}$\\
J0003+2129   &   00:03:19.35   &  21:29:44.40   &  $...$  &  0.45  &  1   &    A   &    $...$  &  $<0.252$   &   $0.035_{-0.000}^{+0.000}$\\
J0004$-$1148  &    00:04:4.92   &   $-$11:48:58.40 &   $...$ &   $...$ &   $...$ &   B  &     $...$ &   $0.350_{-0.059}^{+0.075}$ &   $0.059_{-0.004}^{+0.005}$\\
J0004+2019    &  00:04:35.76   &  20:19:42.20   &  $...$  &  0.677 &  1 &      B  &     $...$ &   $1.139_{-0.353}^{+0.793}$ &   $0.298_{-0.113}^{+0.165}$\\
J0005$-$1648   &   00:05:17.93  &   $-$16:48:4.70  &   $...$ &   $...$ &   $...$ &   U  &     $...$  &  $0.082_{-0.019}^{+0.022}$ &    $0.419_{-0.009}^{+0.009}$\\
J0005+0524   &   00:05:20.21   &  05:24:10.70   &  $...$  &  1.9   &  1  &     Q  &     $...$ &   $<0.044$   &   $1.383_{-0.000}^{+0.000}$ \\
J0005+3820   &   00:05:57.18   &  38:20:15.20   &  2FGLJ0006.1+3821   &     0.229 &  1  &     Q  &     LSP  &  $< 0.105$   &    $0.267_{-0.000}^{+0.000}$ \\
J0006$-$0623  &    00:06:13.89  &   $-$06:23:35.30 &   $...$  &  0.347  & 1  &     B   &    $...$ &   $0.387_{-0.056}^{+0.072}$ &   $0.308_{-0.023}^{+0.023}$\\
J0006+2422    &  00:06:48.79  &   24:22:36.50  &   $...$  &  1.684 &  1  &     Q  &     $...$  &  $<0.100 $  &    $0.130_{-0.000}^{+0.000}$ \\
\hline
\end{tabular}
\end{minipage}
%\medskip
Columns are as follows: (1) OVRO Name; (2) RA; (3) DEC; (4) {\it
  Fermi} association; (5) redshift; (6) redshift reference where 1 is
\cite{healey08}, 2 is \cite{abdo10}, 3 is \cite{ackermann11}, and 4 is \cite{landoni12}; (7) optical classification where Q = FSRQ, B = BL~Lac
object, G = galaxy, A = Active galactic nucleus, N = Narrow line
Seyfert I galaxy, and U = unidentified; (8) SED class from Ackermann et
al. (2011); (9) Intrinsic modulation index and its 1$\sigma$ error; (10) Intrinsic mean flux density and its $1\sigma$ error.
%\end{minipage}
\end{table}
\end{landscape}

\end{document}